\documentclass[aps,pra,twocolumn,showpacs,preprintnumbers,amsmath,amssymb,floatfix,superscriptaddress]{revtex4-2}

\usepackage[utf8]{inputenc}
\usepackage{graphicx,dcolumn,bm,amssymb,amsmath,amsfonts,xcolor,multirow,array}
\usepackage{physics}
\usepackage{subfigure}
\usepackage{verbatim}
\usepackage[
pdfstartview=FitBV,
bookmarks=true,
bookmarksopen=true,
colorlinks=true,
linkbordercolor=blue,
linkcolor=blue,
citecolor=blue,
urlcolor=blue]{hyperref}
\usepackage[capitalize]{cleveref}
\crefname{section}{Sec.}{Secs.}
\crefname{table}{Table}{Tables}
\crefname{figure}{Fig.}{Figs.}
\AddToHook{cmd/appendix/before}{%
    \crefalias{section}{appendix}%
    \crefalias{subsection}{appendix}
}

\begin{document}
\author{A. G. L\"ohr}
\email{Alexander.Loehr@mbi-berlin.de}
\affiliation{Max-Born-Institut, Max-Born-Str. 2A, 12489 Berlin, Germany}
\author{O. Smirnova}
\email{smirnova@mbi-berlin.de}
\affiliation{Max-Born-Institut, Max-Born-Str. 2A, 12489 Berlin, Germany}
\affiliation{Technische Universit\"at Berlin, Str. des 17. Juni 135, 10623 Berlin, Germany}
\author{M. Mirahmadi}
\email{marjan.mirahmadi@mbi-berlin.de}
\affiliation{Max-Born-Institut, Max-Born-Str. 2A, 12489 Berlin, Germany}

\title{Chiral Temporal Structures in Molecular Rotational Dynamics}

\begin{abstract}
We introduce chiral rotational wavepackets: coherent superpositions of a few low-lying rotational states of a molecule for which the expectation value of a molecular axis traces a three-dimensional chiral trajectory in the laboratory frame. The handedness is a dynamical property of the quantum state rather than of the equilibrium structure, and we show that it can be imprinted on linear rotors as well as on spherical and symmetric tops. We derive the action of space inversion, time reversal, and their product on such wavepackets in the Wigner $D^J_{MK}$ basis, and define the rotational enantiomers of a given wavepacket. To quantify their handedness we introduce two derivative-based pseudoscalar measures: a geometrical measure built from the curvature and torsion of the trajectory, which is even under time reversal, and a dynamical measure, which is odd under time reversal. The latter is shown to be proportional to the geodesic curvature of the trajectory projected onto the unit sphere, which links it, via the Gauss--Bonnet theorem, to the solid angle enclosed by the projected trajectory and to the parallel-transport holonomy of a molecule-fixed vector. Finally we show that a nonzero geometrical measure is a sufficient condition for the wavepacket to exhibit true chirality through its orientation trajectory in the sense of Barron.
\end{abstract}
\maketitle

\section{Introduction}\label{sec:intro}
Chirality is usually considered as a structural property: an object is chiral if it cannot be superimposed on its mirror image by any proper rotation. For molecules, chirality is therefore associated with the equilibrium nuclear structure and with parity-odd spectroscopic observables. Chirality can, however, also have a dynamical dimension. A classic example is the coherent tunnelling between the two enantiomeric configurations of a chiral molecule with a low inversion barrier \cite{Hund1927,Quack2002}. Other examples include chiral electronic \cite{chen2024chiral,smeers2026imprinting} and vibrational wavepackets in achiral molecules \cite{PhysRevA.109.012810,tikhonov2022pump} and chiral electronic wavepackets in atoms \cite{Mayer2022,kohnke2026multiphoton}. In contrast to stationary chiral electron states in hydrogen \cite{ordonez2019propensity}, in multi-electron atoms such wavepackets are non-stationary and their chirality may evolve in time, as reflected in photoionization \cite{Mayer2022}.

The notion of \emph{temporal chirality} applies to any time-dependent vector that traces a chiral trajectory in space. It was introduced for structured light fields, electronic currents, and polarization dynamics \cite{smirnova2025new, Ayuso2019,Ayuso2022}. Light whose polarization vector draws a three-dimensional chiral Lissajous figure during the laser period, known as synthetic chiral light \cite{Ayuso2019,Mayer2024}, generates strong enantiosensitive signals through purely electric-dipole interactions \cite{Ayuso2019,Ayuso2022}. This concept has its roots in the microwave domain, where three mutually orthogonal fields  envisioned by Kral and Shapiro \cite{kral2001cyclic} drive enantioselective rotational transitions \cite{Patterson2013}; Ref.~\cite{Ordonez2026} shows how such three-colour schemes can be translated to the optical domain, where interference between two- and three-photon electric-dipole pathways yields enantioselectivities of tens of percent in electronic excitation. Synthetic chiral light has now been realized experimentally in the optical domain yielding chiral photoelectron wave-packets in atoms \cite {kohnke2026multiphoton}, enantiosensitive photoionization and photolysis of chiral molecules \cite{Greenwood2026}, and topologically protected enantiosensitive nonlinear emission from chiral powders \cite{Li2026}. These fields imprint, in a controlled way, relative phases between the multiphoton pathways they drive. As shown below, it is precisely such rotation-irreducible relative phases between rotational components that make a rotational wavepacket truly chiral, so that synthetic chiral light is the natural enabling technology for preparing chiral rotational and, more generally, chiral vibronic wavepackets.

Inspired by this concept, here we introduce chiral rotational wavepackets, which imprint a chiral temporal structure on the orientation of a molecular axis $\boldsymbol{\zeta}$. We show that coherent superpositions of a small number of low-lying rotational states can generate molecular axis expectation values that trace three-dimensional chiral trajectories in the space-fixed frame. The resulting handedness is a property of the quantum state, encoded in the time evolution of rotational observables, and it can be created even in an achiral, e.g. diatomic, molecule.

Chiral rotational wavepackets complement established rotational approaches to sensing molecular chirality, which include microwave three-wave mixing \cite{Patterson2013}, laser-induced enantioselective orientation \cite{Yachmenev2016,Tutunnikov2018}, and optical-centrifuge schemes \cite{Milner2019}. Recently, enantiomer-specific rotational wavepacket dynamics were experimentally observed using time-resolved Coulomb explosion imaging, revealing opposite out-of-plane orientations of structural enantiomers undergoing unidirectional rotation. The study also demonstrated that different angular distributions can yield similar orientation expectation values, highlighting the additional dynamical information contained in the full rotational probability distribution \cite{Mizuse2026}. In those schemes the molecule is chiral and its rotational dynamics is used to distinguish the enantiomers; here the molecule can be achiral, and the chirality is a property of the controlled rotational dynamics itself. We focus on low rotational quantum numbers. The opposite extreme of very high rotational excitation is realized in molecular super-rotors \cite{Karczmarek1999,Korobenko2014}, molecules with $J\sim 10^2$ created in an optical centrifuge. Particularly relevant to the present work are super-rotors of molecules that are achiral in their equilibrium geometry but become chiral in very high rotational states, as ro-vibrational coupling changes the molecular structure \cite{bunker2004chirality,PhysRevLett.121.193201}. More recently, Blech {\it et al.} \cite{Blech2026} proposed generating time-odd rotational chirality in structurally achiral molecules through the orientation of angular momentum in the molecular frame, without requiring laboratory-frame molecular orientation. They also proposed photoelectron circular dichroism as a probe of this handedness. Their approach and our description based on the three-dimensional trajectory of the expectation value of a molecular axis in the laboratory frame provide complementary perspectives on chirality arising from rotational motion.

Chiral rotational wavepackets can be constructed for linear rotors, spherical tops, symmetric tops, and asymmetric tops. In linear rotors and in spherical and symmetric tops the curve drawn by the molecular axis during one rotational period is closed, while in asymmetric tops it is generally not. The asymmetric-top case will be discussed in a separate work; since asymmetric-top eigenstates are expanded in the symmetric-top basis, , making the present results directly applicable to this extension.

Quantitative measures of chirality can focus either on structure or on dynamics. For structures, Harris, Kamien and Lubensky \cite{Harris1999} showed that any chiral measure is a pseudoscalar built from three-point correlations of the distribution, e.g. from its traceless quadrupole, octupole and hexadecapole moments, and that no single such parameter can suffice, since an object can be deformed continuously into its mirror image without ever becoming achiral (the ``rubber-glove'' theorem), so that every measure has blind spots and a family of measures is needed. Pisanty \emph{et al.} \cite{Pisanty2026} recently recast and extended this construction as a family of chiral moments, triple products of the tensorial moments of an arbitrary distribution, that also resolve its radial structure and apply directly to wavefunctions and photoelectron distributions; the Harris--Kamien--Lubensky parameter built from traceless rank-2, 3 and 4 moments is recovered as one member of that family. For dynamics, the relevant measures are those for the trajectories drawn by a time-dependent vector, such as the chiral correlation functions introduced for synthetic chiral light \cite{Ayuso2019,Ayuso2022}. In the present case, this vector is the first orientational moment, or equivalently the rank-1 moment, $\langle\boldsymbol{\zeta}\rangle$ of the rotational wavepacket. Here we focus on the dynamical chirality of the rotational wavepacket and introduce derivative-based pseudoscalar measures for the three-dimensional trajectory drawn by the molecular axis. These are adaptations to the present problem of two classical constructions: the regularized torsion density $\kappa^2\alpha$ of a space curve \cite{Efrati2014} and the triple product of position, velocity and acceleration, which is the continuum limit of the chiral correlation function of Ref.~\cite{Ayuso2019}.

The paper is organized as follows. \Cref{sec:dynamics} defines the orientation trajectory of a rotational wavepacket. \Cref{sec:measures} introduces the two chiral measures, relates one of them to the geodesic curvature of the projected trajectory, and connects the latter to the enclosed solid angle and to a parallel-transport holonomy. \Cref{sec:basis} derives the action of space inversion $\mathcal{P}$, time reversal $\mathcal{T}$, and $\mathcal{PT}$ on the rotational basis, and \cref{sec:enantiomers} uses these results to define rotational enantiomers and to compare their trajectories and chiral measures. In particular, \Cref{sec:truefalse} discusses in which sense chiral rotational wavepackets are truly or falsely chiral according to Barron's classification \cite{Barron1986,Barron2004}. A brief discussion about the projection of molecular axis in the molecular frame is given in \cref{sec:Kdep}. In \cref{sec:conclusions} We discuss the conclusion and prospects.

\section{Orientation trajectory of a rotational wavepacket}\label{sec:dynamics}
We describe molecular rotation by the orientation of a right-handed Cartesian molecular frame $(x,y,z)$ relative to the space-fixed frame $(X,Y,Z)$, both with origin at the nuclear center of mass. The orientation is specified by the Euler angles $\Omega=(\phi,\theta,\chi)$ in the $ZYZ$ convention, with the rotation matrix $R(\Omega)$ given in \cref{app:wignerD}. The rotational wavepacket is expanded in the basis of normalized Wigner $D$-functions of integer $J$ \cite{BrownCarrington2003,Zare1988,Varshalovich1988},
\begin{align}\label{eq:basis}
\langle\Omega|J K M\rangle=\sqrt{\frac{2J+1}{8\pi^2}}\,D^{J\,*}_{MK}(\Omega),
\end{align}
where $-J\leq M\leq J$ is the projection of the total rotational angular momentum $\mathbf{J}$ on the space-fixed $Z$ axis and $-J\leq K\leq J$ its projection on the body-fixed $z$ axis. The complex conjugation in \cref{eq:basis} follows the molecular-spectroscopy convention of Brown and Carrington \cite{BrownCarrington2003}. The freely evolving rotational wavepacket is

\begin{align}\label{eq:wavepacket}
\Psi(\Omega,\tau)=\sum_{JMK} b^J_{MK}\,e^{-iE_{JK}\tau}\,\langle\Omega|J K M\rangle ,
\end{align}
with time-independent complex amplitudes $b^J_{MK}$ and dimensionless eigenvalues $E_{JK}$. In the present work the rotor is a rigid symmetric top, $E_{JK} = J(J+1)+(A/B-1)K^2$, where $A$ and $B$ are the rotational constants and $\tau := B t/\hbar$ is the dimensionless time. Linear rotors and spherical tops are included as the special cases $K=0$ and $A=B$, respectively.

We define the rotational chirality of the system through the time-dependent orientation of the molecular figure axis $\hat{z}$ in the space-fixed frame. Let $\boldsymbol{\zeta}=R^{-1}(\Omega)\hat{z}$ denote the space-fixed components of $\hat{z}$ (see \cref{app:wignerD}). Its expectation value,
\begin{align}\label{oritraj}
\langle\boldsymbol{\zeta}\rangle(\tau) &\equiv \langle\Psi(\Omega,\tau)|\boldsymbol{\zeta}|{\Psi(\Omega,\tau)}\rangle \nonumber \\
&= \big(\langle \zeta_X\rangle(\tau),\langle\zeta_Y\rangle(\tau),\langle\zeta_Z\rangle(\tau)\big),
\end{align}
is the degree of one-dimensional orientation in the space-fixed frame. Using the irreducible spherical components $\zeta_q = D^{1*}_{q0}(\Omega)$, $q\in\{0,\pm 1\}$, the Cartesian components read
\begin{align}\label{zetacomps}
    \langle \zeta_{X}\rangle &= -\sqrt{2}\,\Re\langle D^{1*}_{10}(\Omega)\rangle, \nonumber \\
    \langle \zeta_Y\rangle &= -\sqrt{2}\,\Im\langle D^{1*}_{10}(\Omega)\rangle, \nonumber \\
    \langle \zeta_Z\rangle &= \langle D^{1*}_{00}(\Omega)\rangle .
\end{align}
The explicit expressions in terms of the amplitudes $b^J_{MK}$ are given in \cref{app:expect}. In analogy with synthetic chiral light \cite{Ayuso2019}, the chiral rotational dynamics can be visualized as the tip of the vector $\langle\boldsymbol{\zeta}\rangle(\tau)$ tracing a curve in space. We call this curve the \emph{orientation trajectory},
\begin{align}\label{eq:curve_def}
\Gamma = \{\mathbf{\Gamma}(\tau) \,|\, \tau\in[0,\tau_R]\}, \qquad \mathbf{\Gamma}(\tau):= \langle\boldsymbol{\zeta}\rangle(\tau),
\end{align}
where $\tau_R$ is the period after which the trajectory closes (see below). Since $\mathbf{\Gamma}(\tau)$ is a quantum-mechanical mean orientation rather than the classical position of a unit vector, it lies inside the closed unit ball rather than on the unit sphere:
\begin{align}\label{curve}
    \mathbf{\Gamma}(\tau) =  \|\mathbf{\Gamma}(\tau)\|\,\hat{\mathbf{\Gamma}}(\tau) , \qquad 0\leq  \|\mathbf{\Gamma}(\tau)\|\leq 1 .
\end{align}
The direction $\hat{\mathbf{\Gamma}}(\tau)$ describes the angular motion on the unit sphere and shows in which direction the molecule has a net orientation at each time, while the magnitude $\|\mathbf{\Gamma}(\tau)\|$ is the degree of orientation and describes the radial motion caused by changes in the strength of the net orientation.

All orientation trajectories considered in this work are regular curves,
\begin{align}\label{eq:regular}
    \mathbf{\dot{\Gamma}} := \frac{d\mathbf{\Gamma}}{d\tau}\neq 0
\end{align}
at every point, and they do not pass through the origin, so that $\hat{\mathbf{\Gamma}}(\tau)$ is defined everywhere. Here and below a dot denotes the derivative with respect to $\tau$; $\mathbf{\dot{\Gamma}}(\tau_0)$ is the tangent vector to the curve at time $\tau_0$, i.e., the instantaneous velocity of the mean orientation vector.

\begin{figure}
\centering
\includegraphics[width=\linewidth]{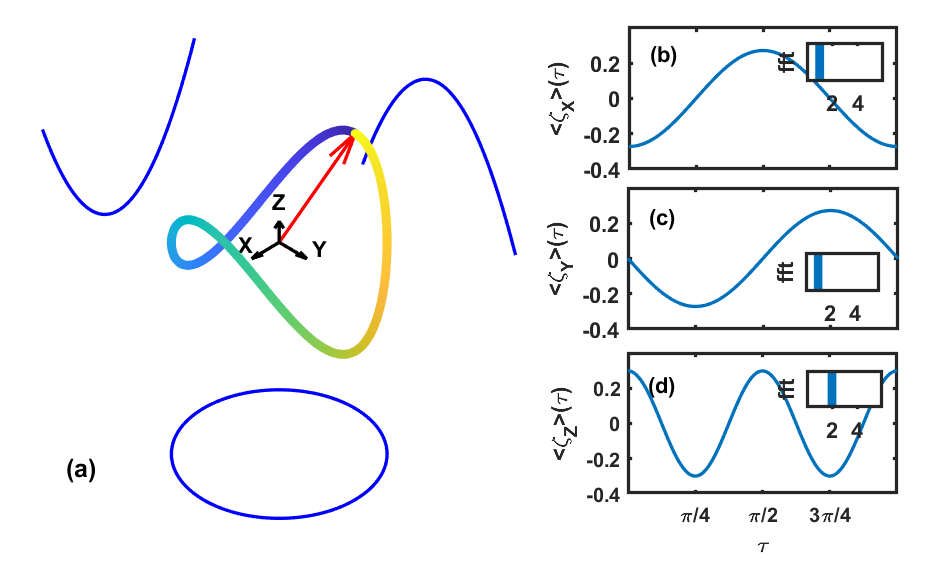}
\caption{\label{fig:potatoR} (a) Orientation trajectory for the initial state $|\psi\rangle^R (\tau=0) = |0\,0\,0\rangle + |1\,0\,1\rangle + |2\,0\,1\rangle$ (notation $|J\,K\,M\rangle$). The color encodes the time $\tau$, from dark blue at $\tau=0$ to yellow at $\tau = \tau_R=\pi$; the red arrow indicates $\langle\boldsymbol\zeta\rangle(\tau= 0)$; and the blue curves are the projections of the trajectory on $XY$, $YZ$ and $XZ$ planes. (b)--(d) Time evolution of the three Cartesian components over one period; the insets show the corresponding Fourier frequencies $\Delta E_{\rm rot}$.}
\includegraphics[width=\linewidth]{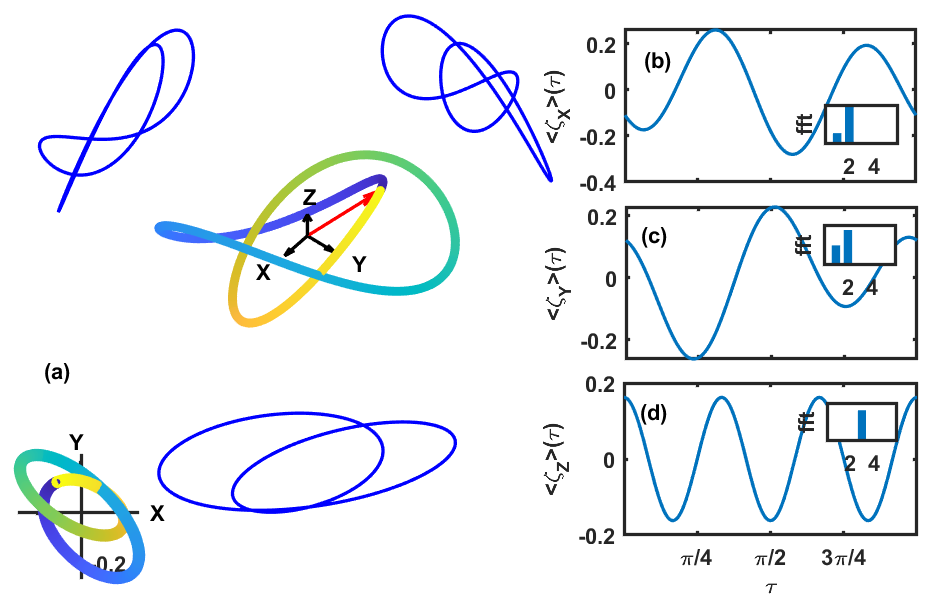}
\caption{\label{fig:InfinityR} (a) Orientation trajectory for the initial state $|\psi\rangle^R(\tau=0) = |0\,0\,0\rangle - e^{2i}|1\,0\,1\rangle + |1\,0\,{-1}\rangle + 3|2\,0\,2\rangle - |2\,0\,{-2}\rangle + |3\,0\,2\rangle $. The inset shows the $XY$ projection of the trajectory, for comparison with the two left-handed partners in \cref{fig:InfinityLref,fig:InfinityLT}. (b)--(d) as in \cref{fig:potatoR}.}
\end{figure}

When each Cartesian component of $\mathbf{\Gamma}(\tau)$ is a finite Fourier series in $\tau$, $\Gamma$ is a so-called Fourier figure, in analogy with the concept of Fourier knots \cite{Kauffman1998}. If each component contains only a single frequency one obtains the special case of a three-dimensional Lissajous curve (see \cref{app:fourier}). \Cref{fig:potatoR} shows a Lissajous-type orientation trajectory together with its three Cartesian components and the frequencies contributing to each of them; \cref{fig:InfinityR} shows a more general Fourier figure. For a rigid symmetric top all frequencies contributing to $\langle\boldsymbol{\zeta}\rangle(\tau)$ are commensurate, $2(J+1)$ with integer $J$, and the trajectory closes after $\tau_R=\pi$ (\cref{app:fourier}). For Fourier figures with arbitrary frequencies the usual commensurability rule applies: if all frequencies are rationally related the curve closes, and if at least one ratio is irrational the curve is quasi-periodic and never closes. 
%In the latter case, relevant for asymmetric tops, the global measures introduced below are to be understood as long-time averages.

A chiral orientation trajectory must be non-planar, i.e., its torsion (\cref{sec:measures}) must be nonzero somewhere along the curve. If the amplitudes $b^J_{MK}$ in \cref{eq:wavepacket} lack the required coherences, e.g. if the wavepacket contains only a single $J$, or only $J$ manifolds that are not coupled by the $\Delta J=\pm 1$ selection rule entering $\langle \zeta_q\rangle$, then $\mathbf{\Gamma}(\tau)$ is constant (possibly zero). A non-planar Fourier figure further requires both longitudinal $(\Delta M=0)$ and transverse $(\Delta M=\pm1)$ coherences with at least two distinct rotational transition frequencies. A minimal population pattern therefore involves three consecutive $J$ manifolds, for example $(J,M)$, $(J+1,M+1)$, and $(J+2,M+1)$ for a given $K$.

\section{Chiral measures}\label{sec:measures}
It is important to distinguish the chirality of the set of points $\Gamma$ from the chirality of the parametrized motion $\mathbf{\Gamma}(\tau)$. The set $\Gamma$ carries no information about the direction in which the curve is traversed, whereas the parametrized trajectory contains both the spatial shape of the curve and its sense of motion. We therefore introduce pseudoscalar measures that respond not only to mirroring but also to the reversal of motion. We emphasize that the vanishing of global measures does not by itself imply that the rotational dynamics are achiral: the trajectory can have substantial local three-dimensional handedness, but zero global measure (more details further below). Nor is the sign of these measures an absolute handedness: different measures of the same object can disagree, because handedness depends on the property being probed \cite{Harris1999,Efrati2014}, as the opposite signs of different measures for the trajectory  in \cref{tab:trajcomp} illustrate. The measures below are therefore chirality \emph{indicators} whose sign is guaranteed to flip under the relevant improper transformations, and that sign change, not the sign itself, is their physically meaningful content.

We use the Frenet (TNB) frame consisting of the unit tangent $\mathbf{T}$, unit normal $\mathbf{N}$, and unit binormal $\mathbf{B}$ as a moving/twisting orthonormal basis along the orientation trajectory. Together with the Frenet--Serret formulas (\cref{app:TNB}), it fully describes the kinematics of a point moving along a differentiable curve in $\mathbb R^3$ \cite{Baer2010,Pressley2010,Crenshaw1993a,Crenshaw1993b,Efrati2014}.

The instantaneous velocity of the mean orientation vector (i.e., at each point in time) is $\mathbf{v}\equiv\dot{\mathbf{\Gamma}}=v\mathbf{T}$, with $v=\|\dot{\mathbf{\Gamma}}\|$. Differentiating once more gives the instantaneous acceleration
\begin{align}
\mathbf{a} \equiv \mathbf{\ddot{\Gamma}}
= \frac{d}{d\tau}\left(v\mathbf{T}\right)
= \dot v\,\mathbf{T} + v^{2}\kappa\,\mathbf{N},
\end{align}
where we have used the Frenet--Serret relation \cref{fs1}. This is the familiar decomposition of the acceleration into a tangential component $\mathbf{a}_{\parallel} := \dot v\mathbf{T}$, which changes the magnitude of the velocity, and a normal component $\mathbf{a}_{\perp} := v^{2}\kappa\mathbf{N}$, which changes its direction. Taking the cross product with $\mathbf{T}$ and using \cref{Tvec} one obtains
\begin{align}\label{eq:kappaB}
\frac{\mathbf{\dot{\Gamma}}\times\mathbf{\ddot{\Gamma}}}{ \|\mathbf{\dot{\Gamma}}\|^{3}} = \frac{\mathbf{v}\times\mathbf{a}}{ v^{3}} = \kappa\mathbf{B}.
\end{align}
The cross product $\mathbf{\dot{\Gamma}}\times\mathbf{\ddot{\Gamma}}$ thus removes the tangential component of the acceleration and keeps only the component responsible for bending the trajectory, with direction along the binormal,
\begin{align}\label{binormal}
\mathbf{B}(\tau)=\frac{\mathbf{\dot{\Gamma}}(\tau)\times \mathbf{\ddot{\Gamma}}(\tau)}{\| \mathbf{\dot{\Gamma}}(\tau)\times \mathbf{\ddot{\Gamma}}(\tau)\|},
\end{align}
and magnitude given by the curvature
\begin{align}\label{tcurvature}
\kappa(\tau)=\frac{\| \mathbf{\dot{\Gamma}}(\tau)\times \mathbf{\ddot{\Gamma}}(\tau)\|}{\| \mathbf{\dot{\Gamma}}(\tau)\|^{3}} .
\end{align}
Under motion reversal $\mathbf{T}\rightarrow -\mathbf{T}$ and $\mathbf{N}\rightarrow \mathbf{N}$, hence $\mathbf{B}\rightarrow -\mathbf{B}$. Under a reflection $\sigma$, $\mathbf{B}$ transforms as a pseudovector, $\mathbf{B}\rightarrow (\det\sigma)\,\sigma\mathbf{B}$.

The curve (in three dimensions) is locally planar up to second order (the osculating plane, \cref{app:TNB}), and $\mathbf{B}$ is the normal to that plane. Scalar products with $\mathbf{B}$ therefore measure signed out-of-plane quantities. In what follows, we introduce two chiral measures based on this.

\subsection{Geometrical measure}\label{sec:geom}
The geometry of the smooth, regular trajectory $\mathbf{\Gamma}$ is characterized by its curvature $\kappa$, \cref{scurvature,tcurvature}, and its torsion $\alpha$,
\begin{align}\label{eq:torsion}
\alpha(\tau)=\frac{\mathbf{B}(\tau)\cdot \mathbf{\dddot{\Gamma}}(\tau)}{\| \mathbf{\dot{\Gamma}}(\tau)\times \mathbf{\ddot{\Gamma}}(\tau)\|}
=\frac{(\mathbf{\dot{\Gamma}}(\tau)\times \mathbf{\ddot{\Gamma}}(\tau))\cdot \mathbf{\dddot{\Gamma}}(\tau)}{\| \mathbf{\dot{\Gamma}}(\tau)\times \mathbf{\ddot{\Gamma}}(\tau)\|^{2}},
\end{align}
which follows from \cref{highdev,binormal,tcurvature}. The torsion is a pseudoscalar and changes sign under reflection of the trajectory, but it is undefined at points of vanishing curvature. Using this fact we define the local chiral measure as $h_g =\kappa^2\alpha$~\cite{Bates2016,Wang2025}, which vanishes for the straight line, while avoiding singularities. Thus we introduce 
\begin{align}\label{gmeasure}
   h_g(\tau) =  \frac{(\mathbf{\dot{\Gamma}}(\tau)\times \mathbf{\ddot{\Gamma}}(\tau))\cdot \mathbf{\dddot{\Gamma}}(\tau)}{\| \mathbf{\dot{\Gamma}}(\tau)\|^{6}},
\end{align}
as the local pseudoscalar measure of the geometrical handedness of the trajectory~\cite{Efrati2014}. Note that $h_g$ is invariant under the reversal of motion, because $\dot{\mathbf{\Gamma}}$ and $\dddot{\mathbf{\Gamma}}$ both change sign while $\ddot{\mathbf{\Gamma}}$ does not. The corresponding global geometrical measure is the arc-length average of $h_g$,
\begin{align}
\mathcal{C}_g &=\frac{1}{L}\int_{0}^{\tau_R} h_g(\tau) \,\|\mathbf{\dot{\Gamma}}(\tau)\|\,d\tau , \label{eq:Cg}\\
L & = \oint_\Gamma ds = \int_{0}^{\tau_R} \|\mathbf{\dot{\Gamma}}(\tau)\|\,d\tau , \label{eq:L}
\end{align}
with $L$ the length of the trajectory, \cref{arclength}. Both $h_g$ and $\mathcal{C}_g$ are invariant under proper rotations and under reparametrization of the curve, odd under spatial reflections and inversion, and even under motion reversal. As a side remark, derivative-based quantities of this type are geometrical and do not, in general, decide whether the trajectory is topologically equivalent to its mirror image.

\subsection{Dynamical measure}\label{sec:dyn}
We introduce the local dynamical chiral measure based on the binormal vector as
\begin{align}\label{dmeasure}
h_d(\tau)&=\kappa \, \mathbf{\Gamma}(\tau)\cdot \mathbf{B}(\tau)
=\dfrac{\mathbf{\Gamma}(\tau)\cdot (\mathbf{\dot\Gamma}(\tau)\times \mathbf{\ddot\Gamma}(\tau))}{\| \mathbf{\dot\Gamma}(\tau)\|^3} .
\end{align}
At each point, $h_d$ is the signed displacement of the trajectory from the origin along the normal vector to the local osculating plane, weighted by the local curvature. It changes sign under reflection and inversion, and, since only $\mathbf{B}$ flips, also under reversal of the direction of motion. Note that with the space-fixed frame origin retained, $h_d$ is not an intrinsic measure of the curve geometry alone, but instead quantifies dynamical handedness with respect to the physically distinguished zero-orientation point $\boldsymbol{\Gamma}=0$ (origin of space-fixed frame). However, one can redefine this measure by replacing $\mathbf{\Gamma}$ with $(\mathbf{\Gamma}(\tau)-\mathbf{\bar{\Gamma}})$ and study chirality only associated with the dynamics of the molecular axis around a constant orientation direction $\mathbf{\bar{\Gamma}}$. This point will become more important in our future works concerning electronic transitions.

We label $h_d$ as a dynamical measure because it distinguishes the two senses of traversal of the same curve; note, however, that it is invariant under reparametrization (both numerator and denominator scale as the cube of a rescaling of $\tau$), so it depends only on the oriented curve and the origin, not on the speed profile along the curve. The global dynamical measure is
\begin{align}\label{eq:Cd}
\mathcal{C}_d &=\frac{1}{L}\int_{0}^{\tau_R} h_d(\tau) \,\|\mathbf{\dot{\Gamma}}(\tau)\|\,d\tau .
\end{align}
For a planar trajectory whose plane contains the origin, $\mathbf{\Gamma}$ lies in the osculating plane and $\mathbf{B}$ is perpendicular to it, so $h_d(\tau)=0$ at every regular point and $\mathcal{C}_d=0$. A planar trajectory whose plane does \emph{not} contain the origin, in contrast, has a nonzero $h_d$ even though its intrinsic geometrical chirality vanishes, $\mathcal{C}_g=0$.
%this is the situation of a rotating polar vector that is not orthogonal to a fixed axis, and we return to its interpretation in \cref{sec:truefalse}. 
Clearly, replacing $\mathbf{\Gamma}(\tau)$ by $\mathbf{\Gamma}(\tau)-\bar{\mathbf{\Gamma}}$ would change this and $h_d$ vanishes everywhere for any type of planar trajectory. 

The dynamical measure is the continuum limit of the chiral correlation function $H^{(3)}$ introduced for synthetic chiral light \cite{Ayuso2019,Ayuso2022}, which is the triple product of the vector at three consecutive times $\tau_0$, $\tau_0+\Delta\tau$, $\tau_0+2\Delta\tau$. As shown in \cref{app:H3},
\begin{align}\label{eq:H3limit}
   \lim_{\Delta\tau\rightarrow 0}\frac{H^{(3)}_{\rm loc}(\tau_0;\Delta\tau)}{(\Delta\tau)^3} = \|\mathbf{\dot\Gamma}(\tau_0)\|^3\, h_d(\tau_0),
\end{align}
so that for a unit-speed parametrization of the trajectory the time average of $H^{(3)}/(\Delta\tau)^3$ coincides with $\mathcal{C}_d$.

\subsubsection{Geometrical meaning of the dynamical measure}\label{sec:geodesic}
Although $h_d$ was introduced as a dynamical measure, it has a clear geometrical meaning. Projecting the orientation trajectory radially onto the unit sphere $S^2$, i.e., normalizing it, removes the variation of the degree of orientation $\|\bm \Gamma\|$ and leaves only the evolution of the direction $\hat{\bm\Gamma}$. As shown in \cref{app:geodesic}, $h_d$ is proportional to the geodesic curvature $\kappa_g$ of the projected trajectory,
\begin{align}\label{eq:hd_kg}
     h_d= \left(\dfrac{\|\mathbf{\dot\Gamma}\|}{\|\mathbf{\Gamma}\| \, \| \dot{\hat{\bm\Gamma}}\|} \right)^{-3} \kappa_g .
\end{align}
Since $\|\dot{\hat{\bm\Gamma}}\|\leq \|\dot{\bm\Gamma}\|/\|\bm\Gamma\|$, the prefactor is at most one: $|h_d|\leq|\kappa_g|$, with equality when the degree of orientation is constant. The radial breathing of the trajectory therefore only damps the local dynamical measure without affecting its sign, which is the sign of $\kappa_g$.

\begin{figure}
\centering
\includegraphics[width=1\linewidth]{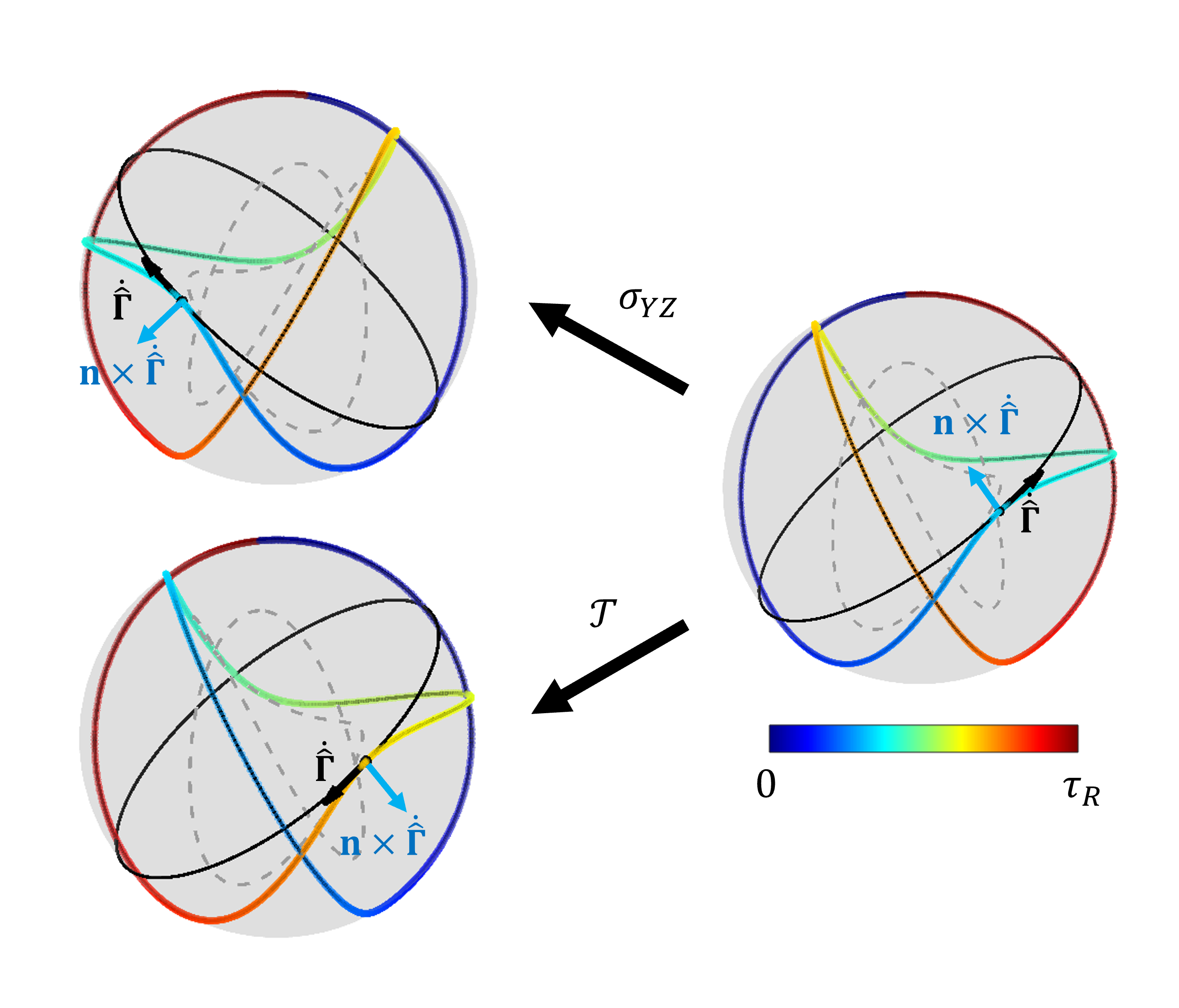}
\caption{\label{fig:geodesics} Effect of reflection $\sigma_{YZ}$ and motion reversal $\mathcal{T}$ on the geodesic curvature of the radial projection of the orientation trajectory onto $S^2$ (gray sphere). The original trajectory is that of \cref{fig:InfinityR}, shown as a gray dashed curve inside the sphere. Under both operations the unit tangent vector $\dot{\hat{\bm \Gamma}}$ (black) and, consequently, $\mathbf{n}\times\dot{\hat{\bm \Gamma}}$ (blue) flip. The black great circle is the geodesic tangent to the trajectory at the chosen point $\tau_0$.}
\end{figure}

The signed geodesic curvature of a curve on a smooth surface quantifies how the curve steers away from the geodesic tangent to it at that point, i.e., the turning of the tangent vector within the surface \cite{Pressley2010}. It changes sign when either the direction of the tangent vector $\dot{\hat{\bm\Gamma}}$ or the surface normal $\mathbf{n}$ is reversed, i.e., when the direction of $\mathbf{n}\times\dot{\hat{\bm \Gamma}}$ flips. \Cref{fig:geodesics} illustrates this for the projected trajectory of \cref{fig:InfinityR} and its reflected and motion-reversed partners. This provides an independent proof that $h_d$ captures both reflection and reversal of motion, and gives it the interpretation of a signed measure of how strongly the mean molecular-axis direction deviates from a great circle on the orientation sphere.

\subsubsection{Solid angle and parallel-transport holonomy}\label{sec:solidangle}
The Gauss--Bonnet theorem relates the integrated geodesic curvature of a simple closed curve on $S^2$ to the solid angle $\Omega_s$ of the region it encloses \cite{Pressley2010},
\begin{align}\label{nsolid}
    \Omega_s = 2\pi - \oint_{\hat{\Gamma}} \kappa_g \, ds
             = 2\pi - \int_0^{\tau_R} \kappa_g(\tau)\,\|\dot{\hat{\bm\Gamma}}(\tau)\| \, d\tau ,
\end{align}
where we have used that the Gaussian curvature of the unit sphere is one and that the area element equals the solid-angle element. In general, however, the projection $\hat{\bm \Gamma}$ has self-crossings, whether or not $\bm \Gamma$ itself does (see \cref{fig:geodesics}). The curve then partitions the sphere into regions $S_i$ bounded by piecewise smooth curves with corners, i.e., curvilinear polygons, for each of which the Gauss--Bonnet theorem reads \cite{Pressley2010}
\begin{align}\label{polygon_GB}
    \Omega_{s_i} = 2\pi - \sum_j\delta_j - \oint_{\partial S_i} \kappa_g \, ds ,
\end{align}
where $\delta_j$ is the exterior angle at the $j$-th vertex. Resolving each transversal (X-type) crossing into two non-crossing arcs decomposes the curve into simple closed loops. Applying \cref{polygon_GB} to each loop and summing, the exterior-angle terms cancel pairwise and one obtains \cite{Binysh2018}
\begin{align}\label{nsolid_sc}
    \Omega_s = 2\pi(m+1) - \oint_{\hat{\Gamma}} \kappa_g \, ds \pmod{4\pi},
\end{align}
where $m$ is the number of crossings. The solid angle is defined only modulo $4\pi$, because two spanning surfaces of the same oriented curve can differ by the entire sphere; we choose the representative $\Omega_s\in[-2\pi,2\pi)$. \Cref{nsolid_sc} requires all crossings to be transversal, i.e., the curve intersects itself at each crossing exactly once rather than touching itself.

In this way the local geodesic curvature of the projected trajectory is connected to the global solid angle it encloses. The same geometry governs the parallel transport of vectors along the projected trajectory. Consider a vector $\mathbf{X}$ tangent to $S^2$ that is parallel transported along the closed curve $\hat{\bm\Gamma}$; after one traversal it returns rotated with respect to its initial direction by the holonomy angle $\Phi_{\rm hol}$. As shown in \cref{app:geodesic}, $\Phi_{\rm hol}=-\oint\kappa_g\,ds \pmod{2\pi}$, so that
\begin{align}\label{eq:holonomy}
    \Phi_{\rm hol}\,[\text{rad}] \equiv \Omega_s\,[\text{sr}] \pmod{2\pi},
\end{align}
with $\Phi_{\rm hol}\in[-\pi,\pi)$; the identification of radians with steradians is legitimate on the unit sphere. We stress that this is a geometric construction: the holonomy characterizes the oriented spherical motion of the mean molecular axis, and only the direction of the orientation contributes to it, not its degree. Whether it manifests in a physical observable, e.g. through a molecule-fixed transition dipole in a pump--probe scheme, is left for future work.

The quantities $\mathcal{C}_d$, $\Omega_s$ and $\Phi_{\rm hol}$ all vanish, through cancellation of opposite local contributions, for trajectories whose projection encloses no net signed solid angle, while $\mathcal{C}_g$ may remain finite; a trefoil-shaped trajectory is an example. Conversely, for a trajectory with the symmetry of a saddle (the "potato chip" of \cref{fig:trajs,fig:potatoR}) all global quantities vanish, including $\mathcal{C}_g$, consistent with the absence of a net handedness of the curve itself. Global measures thus distinguish the three-dimensional handed geometry of the curve from its net oriented spherical motion.

\section{Rotational basis and its properties under space inversion and motion reversal}\label{sec:basis}
To define the enantiomeric partner of a chiral rotational wavepacket we first derive how the basis functions transform under space inversion (parity, $\mathcal{P}$), which connects the two enantiomers of a chiral molecule, and under motion reversal (time reversal, $\mathcal{T}$).

\subsection{Space inversion}\label{sec:parity}
\begin{figure}
\centering
\includegraphics[width=\linewidth]{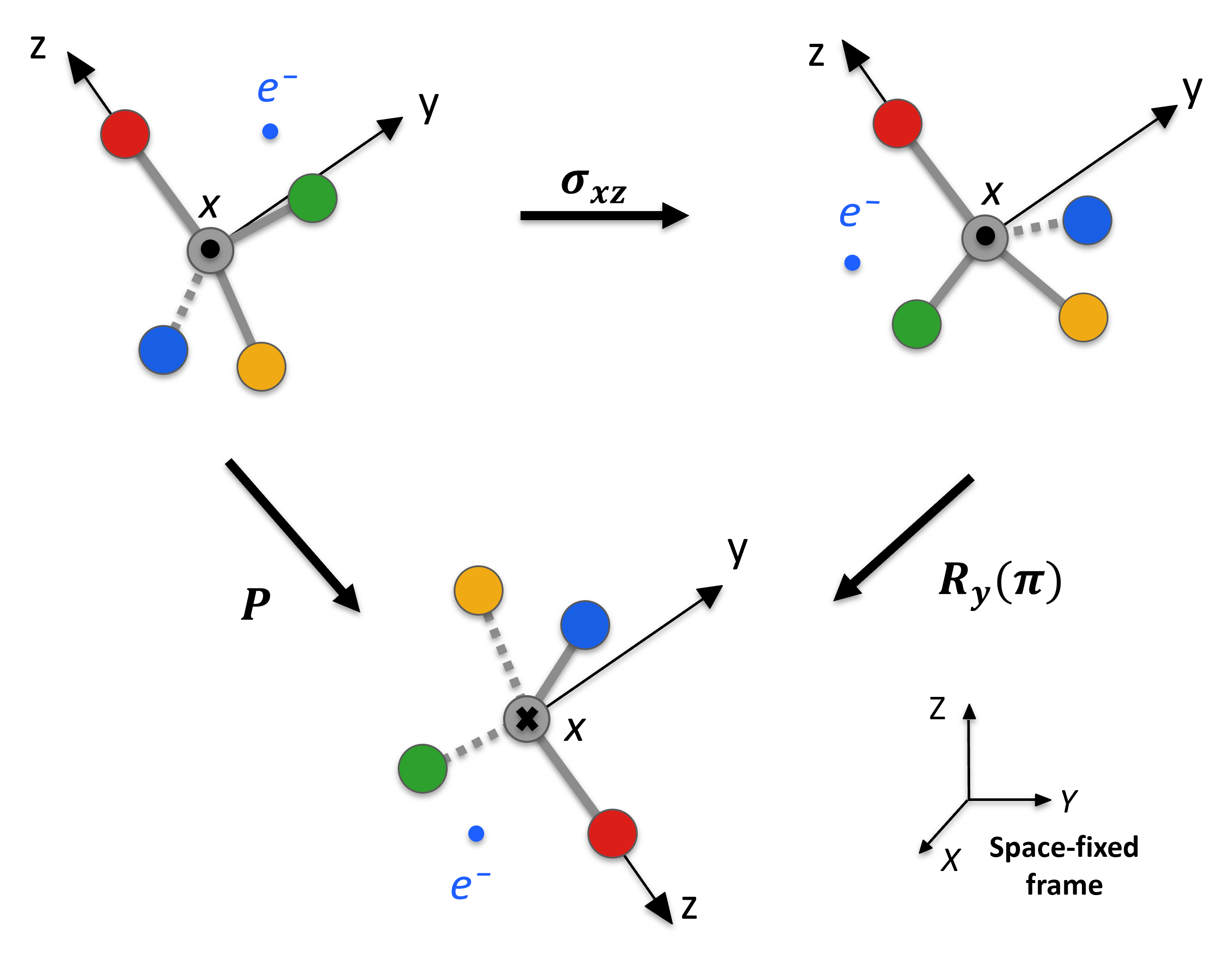}
\caption{\label{fig:parity} Space-fixed inversion $\mathcal{P}$ of a molecule, decomposed into a reflection $\sigma_{xz}$ in the molecular frame followed by a rotation of the whole structure, including the body-fixed axes, by $\pi$ about $y$. After inversion the body-fixed axes are reassigned such that $z$ and $x$ are defined by the same atoms as before, while $y$ is chosen to keep the frame right-handed.}
\end{figure}

Spatial inversion maps the nuclear and electronic configuration of one enantiomer onto the other while leaving angular momenta unchanged, $\mathcal{P}\mathbf{J}\mathcal{P}^{-1} = \mathbf{J}$. For molecular rotations it is necessary to distinguish the molecule-fixed inversion, which leads to the $g/u$ classification of states, from the space-fixed inversion $\mathcal{P}$; the two are not equivalent. Here we consider the space-fixed inversion acting on space-fixed coordinates,
\begin{align}
    \mathcal{P}: (X,Y,Z) \rightarrow (-X,-Y,-Z).
\end{align}
Following Hougen \cite{Hougen1970} and Larsson \cite{Larsson1981} (see also Ref.~\cite{BrownCarrington2003}), and keeping the molecular frame right-handed, $\mathcal{P}$ is equivalent to a reflection in a molecule-fixed plane (here the $xz$ plane) followed by a rotation by $\pi$ about the molecular $y$ axis, as illustrated in \cref{fig:parity}. The reflection acts on the internal (vibrational, electronic, and spin) coordinates, while the rotation acts on the Euler angles as
\begin{align}\label{eq:Peuler}
    \phi \rightarrow \phi+\pi, \qquad
    \theta \rightarrow \pi-\theta, \qquad
    \chi \rightarrow \pi-\chi .
\end{align}
The effect of $\mathcal{P}$ on the basis functions is therefore
\begin{align}\label{PD}
    \mathcal{P} D^{J*}_{M K}(\phi,\theta,\chi) &\equiv R_y(\pi)D^{J*}_{M K}(\phi,\theta,\chi) \nonumber \\
    & = D^{J*}_{M K}(\pi+\phi,\pi-\theta,\pi-\chi) \nonumber \\
    & = (-1)^{J-K}D^{J*}_{M\, -K}(\phi,\theta,\chi),
\end{align}
i.e.,
\begin{align}\label{eq:Pket}
    \mathcal{P}|J K M\rangle = (-1)^{J-K}|J\,{-K}\, M\rangle .
\end{align}
The projection $M$ on the space-fixed $Z$ axis is unchanged, while the sign of $K$ flips: although $\mathbf{J}$ is even under parity, the body-fixed $z$ axis is mapped to $-z$ when the inverted molecule is reassigned a right-handed frame. In the $\{|JKM\rangle\}$ basis, inversion therefore acts as a (convention-dependent) transformation of the $K$ components into $-K$ components. This subtlety is irrelevant for linear rotors ($K=0$) but remains relevant for symmetric tops, and becomes more complicated for asymmetric tops, where $K$ is not a good quantum number and the rotational eigenstates are superpositions of symmetric top states. 

\subsection{Reversal of motion}\label{sec:timereversal}
The time-reversal operator $\mathcal{T}$ is antiunitary and acts on a quantum state as \cite{Wigner1959,Sakurai2017}
\begin{align}
   \mathcal{T}\Psi(\Omega,t) = \Psi^*(\Omega,-t).
\end{align}
This is not the application of an operator at a fixed time; it produces a state whose forward evolution reproduces the original motion backward,
\begin{align}
    \Psi(\Omega,t) &= \sum_{J}\psi_{J}(\Omega)e^{-iE_Jt/\hbar} \nonumber \\
    &\xrightarrow{\;\mathcal{T}\;}  \Psi^*(\Omega,-t) = \sum_{J}\psi_{J}^*(\Omega)e^{-iE_Jt/\hbar} .
\end{align}
Following Wigner, it is therefore more appropriate to call $\mathcal{T}$ the \emph{motion-reversal} operator. Consistently, the angular momentum is odd under $\mathcal{T}$,
\begin{align}\label{TJ}
    \mathcal{T}\mathbf{J}\mathcal{T}^{-1} = -\mathbf{J},
\end{align}
so that $\mathcal{T}$ reverses the sign of all angular-momentum projections, i.e., the sense of rotation. Since the Euler angles are even under time reversal, the action of $\mathcal{T}$ on the basis functions is given by complex conjugation alone,

\begin{align}\label{TD}
    \mathcal{T} D^{J*}_{M K}(\phi,\theta,\chi) &\equiv  D^{J}_{M K}(\phi,\theta,\chi) \nonumber \\
    &= (-1)^{M-K} D^{J*}_{-M\, -K}(\phi,\theta,\chi),
\end{align}
where the second equality is a symmetry property of the Wigner $D$-functions \cite{Varshalovich1988}. Hence

\begin{align}\label{eq:Tket}
    \mathcal{T}|J K M\rangle = (-1)^{M-K}|J\,{-K}\,{-M}\rangle .
\end{align}
On this basis, the effect of $\mathcal{T}$ is thus equivalent to a space-fixed inversion followed by a rotation by $\pi$ about the space-fixed $Y$ axis.

\subsection{Inversion followed by motion reversal}\label{sec:PT}
It follows that under the combined operation $\mathcal{PT}$ the basis is acted upon only by the rotation $R_Y(\pi)$ about the space-fixed $Y$ axis,
\begin{align}\label{PTD}
    \mathcal{PT}D^{J*}_{MK}(\phi,\theta,\chi) = (-1)^{J-M} D^{J*}_{-M\, K}(\phi,\theta,\chi),
\end{align}
and
\begin{align}\label{eq:PTket}
    \mathcal{PT}|J K M\rangle = (-1)^{J-M}|J\,K\,{-M}\rangle .
\end{align}

\section{Rotational enantiomers}\label{sec:enantiomers}
To identify the enantiomeric partner of the wavepacket of \cref{eq:wavepacket} we compare the action of $\mathcal{P}$, $\mathcal{T}$, and $\mathcal{PT}$ on it. Under space inversion, using \cref{eq:Pket},
\begin{align}\label{Ppsi}
   \mathcal{P}\Psi(\Omega,\tau) &= \sum_{JMK}b^J_{MK}(-1)^{J-K}e^{-iE_{JK}\tau}\langle\Omega|J\,{-K}\,M\rangle \nonumber \\
& = \sum_{JMK}\tilde{b}^J_{MK}\,e^{-iE_{JK}\tau}\langle\Omega|J K M\rangle ,
\end{align}
with $\tilde{b}^J_{MK}:= (-1)^{J+K}b^J_{M\,-K}$. The motion-reversed wavepacket is

\begin{align}\label{Tpsi}
     \mathcal{T}\Psi(\Omega,\tau) &= \sum_{JMK}b^{J*}_{MK}(-1)^{M-K}e^{-iE_{JK}\tau}\langle\Omega|J\,{-K}\,{-M}\rangle \nonumber \\
& = \sum_{JMK}\bar{b}^J_{MK}\,e^{-iE_{JK}\tau}\langle\Omega|J K M\rangle ,
\end{align}
with $\bar{b}^J_{MK}:= (-1)^{K-M}b^{J*}_{-M,-K}$, and the $\mathcal{PT}$-transformed wavepacket is
\begin{align}\label{TPpsi}
     \mathcal{PT}\Psi(\Omega,\tau) &= \sum_{JMK}b^{J*}_{MK}(-1)^{J+M}e^{-iE_{JK}\tau}\langle\Omega|J\,K\,{-M}\rangle \nonumber \\
& = \sum_{JMK}\bar{\bar{b}}^J_{MK}\,e^{-iE_{JK}\tau}\langle\Omega|J K M\rangle ,
\end{align}
with $\bar{\bar{b}}^J_{MK} = (-1)^{J+M} b^{J*}_{-M\, K}$. The phases in \cref{PD,TD,PTD} deserve a comment: for a single basis state they are global phases without observable consequence, but for a wavepacket the relative phases between components are physical, and a wavepacket constructed with or without them is generally different. Equipped with these relations we now compare the orientation trajectories of a wavepacket and of its transformed partners, and use them to define the right- and left-handed rotational enantiomers $|\psi\rangle^R$ and $|\psi\rangle^L$. To simplify our analysis, we will omit the summation over $K$ in \cref{eq:wavepacket,Tpsi} and set it as a constant $K=K_0$. 

\begin{comment}
\begin{table}
\centering
\caption{\label{tab:effects} Effect of transformations of the expansion coefficients on the orientation trajectory $\mathbf{\Gamma}(\tau)$ for a wavepacket with a single value of $K$. The transformations in the last three rows are applied without the phase factors and complex conjugation of \cref{Ppsi,Tpsi,TPpsi}.}
\setlength{\tabcolsep}{5pt}
\renewcommand{\arraystretch}{1.15}
\begin{tabular}{ll}
\hline \hline
Transformation & Effect on the trajectory  \\
\hline
$\mathcal{P}$ & Inversion \\
$\mathcal{T}$ & Motion reversal \\
$\mathcal{PT}$ & Inversion and motion reversal  \\
$K \to -K$ & Inversion of the static offset only\footnote{The time-independent orientation offset, \cref{B4,B10}, changes sign; the oscillating part is unchanged.} \\
$M,K \to -M , -K$ & Reflection $\sigma_{YZ}$ of the directed trajectory \\
$M \to -M $ & Reflection $\sigma_{YZ}$  \\
\hline\hline
\end{tabular}
\end{table}
\end{comment}

The effect of different operators defined above on the trajectory given by~\cref{oritraj} is summarized in \cref{tab:effects}. However, to avoid the complexity, throughout this work we concentrate on the case where both projection numbers $M$ and $K$ change sign:  (i) with phase $(-1)^{K-M}$ which is the standard procedure of motion-reversal (or time-reversal) following \cref{Tpsi}, i.e.,  $\bar{b}^J_{MK}:= (-1)^{K-M}b^{J*}_{-M-K}$, and, (ii) without this phase and conjugation, i.e., replacing the coefficients $\bar{b}^J_{MK}:= b^{J}_{-M-K}$ in \cref{eq:wavepacket}. we focus on the spatial reflection and motion reversal of the orientation trajectory. Using reflection rather than inversion does not restrict the chirality classification, since the $YZ$ reflection and inversion differ only by the proper rotation. 

\begin{table*}[t]
\centering
\caption{\label{tab:effects}
Effect of transformations of the expansion coefficients on the orientation
trajectory $\mathbf{\Gamma}(\tau)$ for a wavepacket with a single value of $K$.
The transformations in the last three rows are applied without the phase
factors and complex conjugation of \cref{Ppsi,Tpsi,TPpsi}.}

\renewcommand{\arraystretch}{1.3}

\begin{tabular*}{\textwidth}{@{\extracolsep{\fill}}ccl}
\hline\hline
Transformation
& Transformation of $\mathbf{\Gamma}(\tau)$
& Geometrical effect \\
\hline

$\mathcal{P}$
& $-\mathbf{\Gamma}(\tau)$
& Inversion \\

$\mathcal{T}$
& $\mathbf{\Gamma}(\tau_R-\tau)$
& Reversal of motion \\

$\mathcal{PT}$
& $-\mathbf{\Gamma}(\tau_R-\tau)$
& Inversion and reversal of motion \\

$K\to-K$
& $-\mathbf{\Gamma}_{\rm stat}+\mathbf{\Gamma}_{\rm osc}(\tau)$
& Inversion of the static offset only\footnote{The time-independent orientation offset, \cref{B4,B10}, changes sign; the oscillating part is unchanged.} \\

$M,K\to-M,-K$
& $\sigma_{YZ}\mathbf{\Gamma}(\tau)$
& Reflection $\sigma_{YZ}$\footnote{The direction of is spatially reflected as well.} \\

$M\to-M$
& $\sigma_{YZ}
[-\mathbf{\Gamma}_{\rm stat}+\mathbf{\Gamma}_{\rm osc}(\tau)]$
& Reflection $\sigma_{YZ}$ and inversion of the static offset \\

\hline\hline
\end{tabular*}
\end{table*}

\subsection{Orientation trajectories}\label{sec:trajectories}

\begin{figure}
\centering
\includegraphics[width=\linewidth]{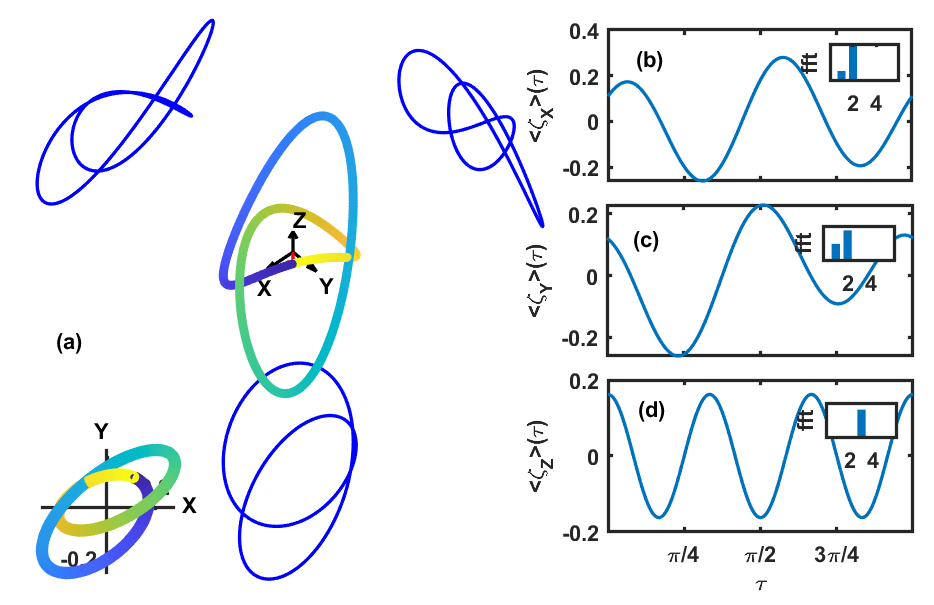}
\caption{\label{fig:InfinityLref} Left-handed partner obtained by reflection $\sigma_{YZ}$ of the trajectory of \cref{fig:InfinityR}, corresponding to $|\psi\rangle^L(\tau=0) = |0\,0\,0\rangle +|1\,0\,1\rangle  - e^{2i}|1\,0\,{-1}\rangle -|2\,0\,2\rangle + 3|2\,0\,{-2}\rangle + |3\,0\,2\rangle $. Details as in \cref{fig:InfinityR}; the inset ($XY$ projection) makes the reflection with respect to \cref{fig:InfinityR} evident.}
\includegraphics[width=\linewidth]{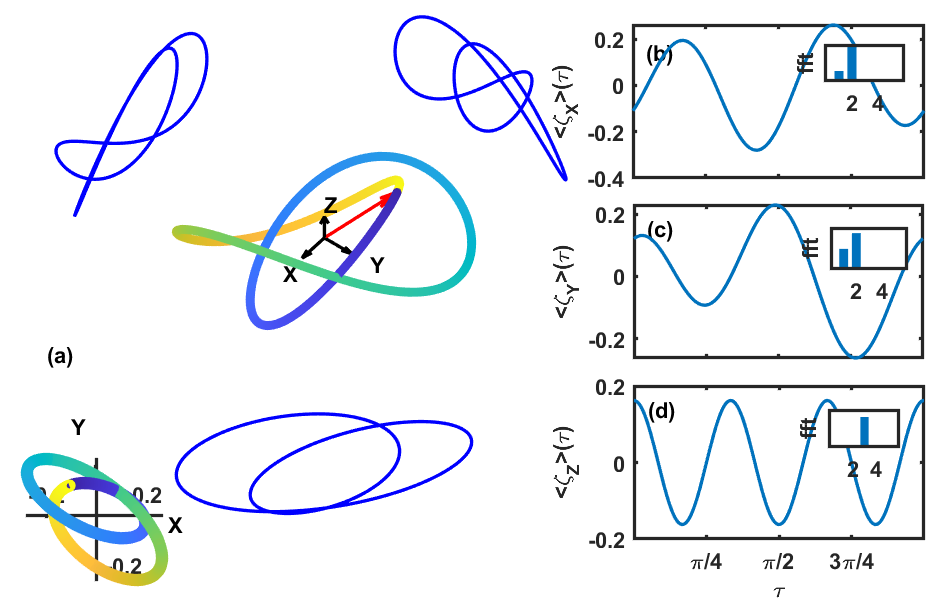}
\caption{\label{fig:InfinityLT} Left-handed partner obtained by motion reversal, \cref{Tpsi}, of the trajectory of \cref{fig:InfinityR}, corresponding to $|\psi\rangle^L(\tau=0) = |0\,0\,0\rangle -|1\,0\,1\rangle  + e^{-2i}|1\,0\,{-1}\rangle -|2\,0\,2\rangle + 3|2\,0\,{-2}\rangle + |3\,0\,2\rangle $. Details as in \cref{fig:InfinityR}.}
\end{figure}

As an example, left enantiomers defined by the two cases mentioned above, associated with the right enantiomer shown in \cref{fig:InfinityR} are depicted in \cref{fig:InfinityLref,fig:InfinityLT}. \Cref{fig:InfinityLref} is a spatial reflection both in the shape of trajectory and the direction of motion (spatial mirror image at the same time $\tau$). Geometrical and dynamical quantities for these trajectories are listed in \cref{tab:trajcomp}.

\begin{table}[b]
\centering
\caption{Geometrical and dynamical quantities for the right-handed trajectory
in \cref{fig:InfinityR} and its spatially reflected counterpart $L_{\mathrm{Ref}}$, and its time-reversed counterpart $L_T$. In this case the number of self crossings in \cref{polygon_GB} is $m=3$.}
\label{tab:trajcomp}
\setlength{\tabcolsep}{16pt}
\renewcommand{\arraystretch}{1.15}
\begin{tabular}{lccc}
\hline\hline

& $R$
& $L_{\mathrm{Ref}}$
& $L_T$ \\
\hline

$\mathcal{C}_g $
& $+9.99$
& $-9.99$
& $+9.99$ \\

$\mathcal{C}_d $
& $+0.09$
& $-0.09$
& $-0.09$ \\

$\Omega_{s}$ (sr)
& $-1.52$
& $+1.52$
& $+1.52$ \\

$\Phi_{\mathrm{hol}}$ (rad)
& $-1.52$
& $+1.52$
& $+1.52$ \\
\hline\hline
\end{tabular}%
\end{table}

\subsection{Dynamical versus Geometrical chiral measures}\label{sec:compare}
\begin{figure}
\centering
\includegraphics[width=0.75\linewidth]{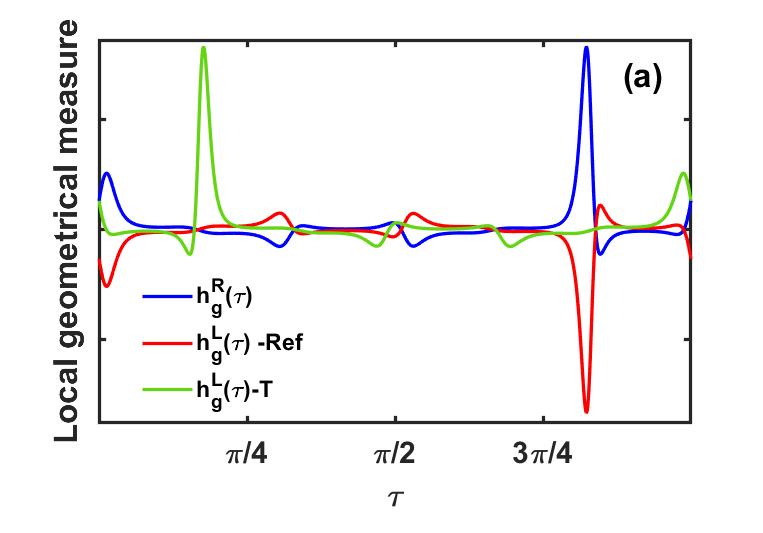}
\includegraphics[width=0.75\linewidth]{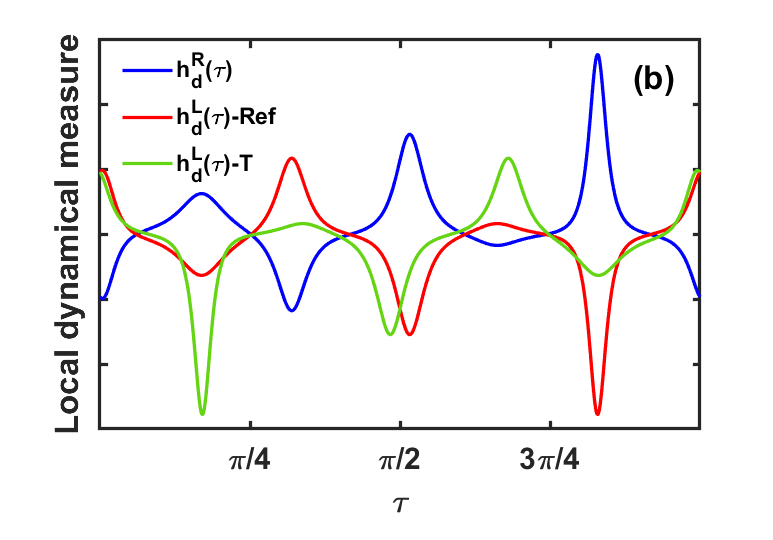}
\caption{\label{fig:localmeasures} Local (a) geometrical and (b) dynamical chiral measures as functions of $\tau$ for the right-handed trajectory of \cref{fig:InfinityR} (blue) and for the two left-handed partners obtained by reflection (red) and by reversal of motion (green).}
\end{figure}

\Cref{fig:localmeasures} shows both local measures for the trajectory of \cref{fig:InfinityR} and for its two partners obtained by reflection and by motion reversal shown in \cref{fig:InfinityLref,fig:InfinityLT}. To see the effect of motion reversal the local measures must be compared at corresponding points of the original and reversed trajectories, i.e., at $\tau$ and $\tau_R-\tau$. Under $\mathcal{T}$ the geometrical measure is unchanged, $h_g^{L}(\tau_R-\tau)=h_g^R(\tau)$, whereas the dynamical measure changes sign, $h_{d}^{L}(\tau_R-\tau)=-h_{d}^R(\tau)$. Under reflection both change sign. Thus $\mathcal{C}_g$ characterizes the three-dimensional handed shape of the curve, while $\mathcal{C}_d$ additionally distinguishes opposite senses of traversal relative to the physical origin. The two measures are complementary; their different behavior under motion reversal is what makes them respectively sensitive to true and to false chirality in the sense of Barron (see below).

\subsection{True and false chirality}\label{sec:truefalse}
Barron \cite{Barron1986,Barron2004} distinguishes true chirality,``possessed by systems that exist in two distinct enantiomeric states that are interconverted by space inversion but not by time reversal combined with any proper spatial rotation", from false chirality, where the two states are also interconverted by time reversal combined with a proper rotation. A stationary spinning cone is the classic example of false chirality; a translating spinning cone is truly chiral. It should be stressed, however, that \emph{false}  does not mean \emph{absent}. The distinction is physically important because only a truly chiral influence can induce an enantiomeric excess at thermodynamic equilibrium, whereas a falsely chiral influence may produce enantioselection under kinetic, nonequilibrium conditions~\cite{Barron1987,Barron1986,Barron2004}. A falsely chiral wavepacket is still
not superimposable on its mirror image by any proper rotation if direction of time is taken into account. Here we deliberately consider only spatial reflection, rather than inversion, since our analysis focuses on the reflected rotational enantiomer. This does not restrict the chirality classification, because the $YZ$ reflection and spatial inversion differ only by a proper rotation by $\pi$ about the $X$ axis. The discussion can therefore be generalized straightforwardly to the inversion-based formulation used in Refs.~\cite{Barron1986,Barron2004}.

For a falsely chiral trajectory in Barron’s sense, the reflected motion must be superimposable on the time-reversed original trajectory by a proper rotation, which for the local geometrical chirality implies

\begin{align}
    h_g^{L-\mathrm{Ref}}(\tau) = h_g^{L-\mathcal{T}}(\tau) \equiv h_g^{R}(\tau_R-\tau).
\end{align}
Since reflection changes the sign of $h_g$, $h_g^R$ is antisymmetric about the midpoint of the period, $h_g^{R}(\tau_R-\tau) = -h_g^R(\tau)$, so the positive and negative local contributions cancel over one complete period. Consequently, the global geometrical chiral measure $\mathcal{C}_{g}$ vanishes for a trajectory with false chirality. However, the converse is not guaranteed:
$\mathcal{C}_{g}=0$ does not by itself prove false chirality, because the integral could vanish accidentally. 
The two measures of \cref{sec:measures} thus have complementary roles in Barron's classification. $\mathcal{C}_g$ is a time-even pseudoscalar, and its nonzero value provides a sufficient indicator of true chirality; $\mathcal{C}_d$, $\Omega_s$ and $\Phi_{\rm hol}$ are time-odd pseudoscalars, and can remain nonzero for \emph{false} chirality that arises from the combination of orientation and rotation with respect to the laboratory frame, although they do not by themselves distinguish false from true chirality. %A time-odd pseudoscalar cannot distinguish the $\mathcal{P}$-partner from the $\mathcal{T}$-partner of a wavepacket (\cref{tab:trajcomp}), which is precisely why it is sensitive to both. 
It is worth mentioning that the geometrical true or false chirality criterion itself, does not rely on the symmetric top approximation and is more general.

A sufficient realization of the false chirality condition, $\langle\boldsymbol{\zeta}\rangle^{L-{\rm Ref}}(\tau) = \langle R^{-1}(\Omega_0)\,\boldsymbol{\zeta}\,R(\Omega_0)\rangle^{L-\mathcal{T}}(\tau) $, for the rotational wavepackets is 
\begin{align}\label{fchi_wf}
   \Psi^{L-{\rm Ref}}(\Omega,\tau) = e^{i\eta} R(\Omega_0)  \Psi^{L-\mathcal{T}}(\Omega,\tau),
\end{align}
where $R(\Omega_0)$ is a proper rotation defined in the space-fixed frame and $e^{i\eta}$ is the overall relative phase between the two wavepackets. 
At the level of the coefficients given by conditions (i) and (ii) in \cref{sec:enantiomers}, \cref{fchi_wf} gives
\begin{align}\label{fchi_coeff}
   b^{J}_{-M-K} = e^{i\eta} \sum_{M''}(-1)^{K-M''}\,D^{J}_{M\,M''}(\Omega_0) \,b^{J*}_{-M''\,-K}~, 
\end{align}
which must hold for every J,M,K component. Since the $\sigma_{YZ}$ reflection (in space-fixed frame) differs from inversion by the proper rotation $R_X(\pi)$, the corresponding coefficient criterion for the inversion-related state is obtained simply by including the additional rotation in \cref{fchi_coeff}. \Cref{fchi_coeff} gives the general criterion for false chirality of the complete rotational wavepacket and is stronger than the criterion based on the first-moment trajectory $\langle\boldsymbol{\zeta}\rangle$. It is possible that the the trajectory is falsely chiral, while some higher-rank orientational moment does not obey the corresponding rotation relation. In this case \cref{fchi_wf} is not satisfied.

For the particular case in which $R$ is a rotation about $Z$ axis by angle $\varphi_0$ (i.e., Wigner $D$-matrix is diagonal in $M$) false chirality requires the coefficient phases to depend linearly on $M$, i.e., $\arg(b^{J}_{-M-K}) = \beta M+\gamma_{(-K)}$ modulo $\pi$, with $\gamma_{(-K)}$ being the $K$-dependent phase offset, such that
\begin{align}\label{fchi_phase}
  2\beta   = \pi- \varphi_0 \pmod{2\pi} \nonumber \\
   2\gamma_{(-K)}  = K\pi + \eta  \pmod{2\pi}  
\end{align}
As a simple limiting case, if the expansion coefficients are real, the phase conditions reduce to $\varphi_0=\pi$ and $\eta=-K\pi$. Hence, for a wavepacket with single $K$, real coefficients necessarily correspond to the false chirality under a $\pi$ rotation about the $Z$ axis.

Although the analytical derivation is formulated in the symmetric-top basis, it is expected to remain applicable to near-oblate or near-prolate chiral molecules, provided that asymmetry induced $K$-mixing is sufficiently weak. In this regime, the conditions \cref{fchi_coeff,fchi_phase} should also remain approximately valid. 

%(ii) The geometrical measure provides a sufficient condition for \emph{true} chirality. $\mathcal{C}_g$ is invariant under proper rotations, reparametrizations and time shifts, odd under $\mathcal{P}$ and even under $\mathcal{T}$ (\cref{sec:geom}). Hence, if $\mathcal{C}_g(\Psi)\neq 0$, then $\mathcal{C}_g(\mathcal{P}\Psi)=-\mathcal{C}_g(\Psi)$ while $\mathcal{C}_g(R\,\mathcal{T}\Psi)=+\mathcal{C}_g(\Psi)$ for every rotation $R$, so that $\mathcal{P}\Psi$ can never be a rotated copy of $\mathcal{T}\Psi$: the wavepacket is truly chiral. Conversely, every falsely chiral wavepacket, and in particular every wavepacket of type (i), has $\mathcal{C}_g=0$, as we found for the saddle-shaped trajectories of \cref{fig:trajs}. The wavepacket of \cref{fig:InfinityR} has $\mathcal{C}_g=+9.99$ (\cref{tab:trajcomp}); it is truly chiral, and indeed its relative phases $-e^{2i}$ on $|1\,0\,1\rangle$ and $+1$ on $|1\,0\,{-1}\rangle$ cannot be removed by any rotation about $Z$.

\subsubsection{A falsely chiral example}\label{sec:potato}
\begin{figure}
\includegraphics[width=0.95\linewidth]{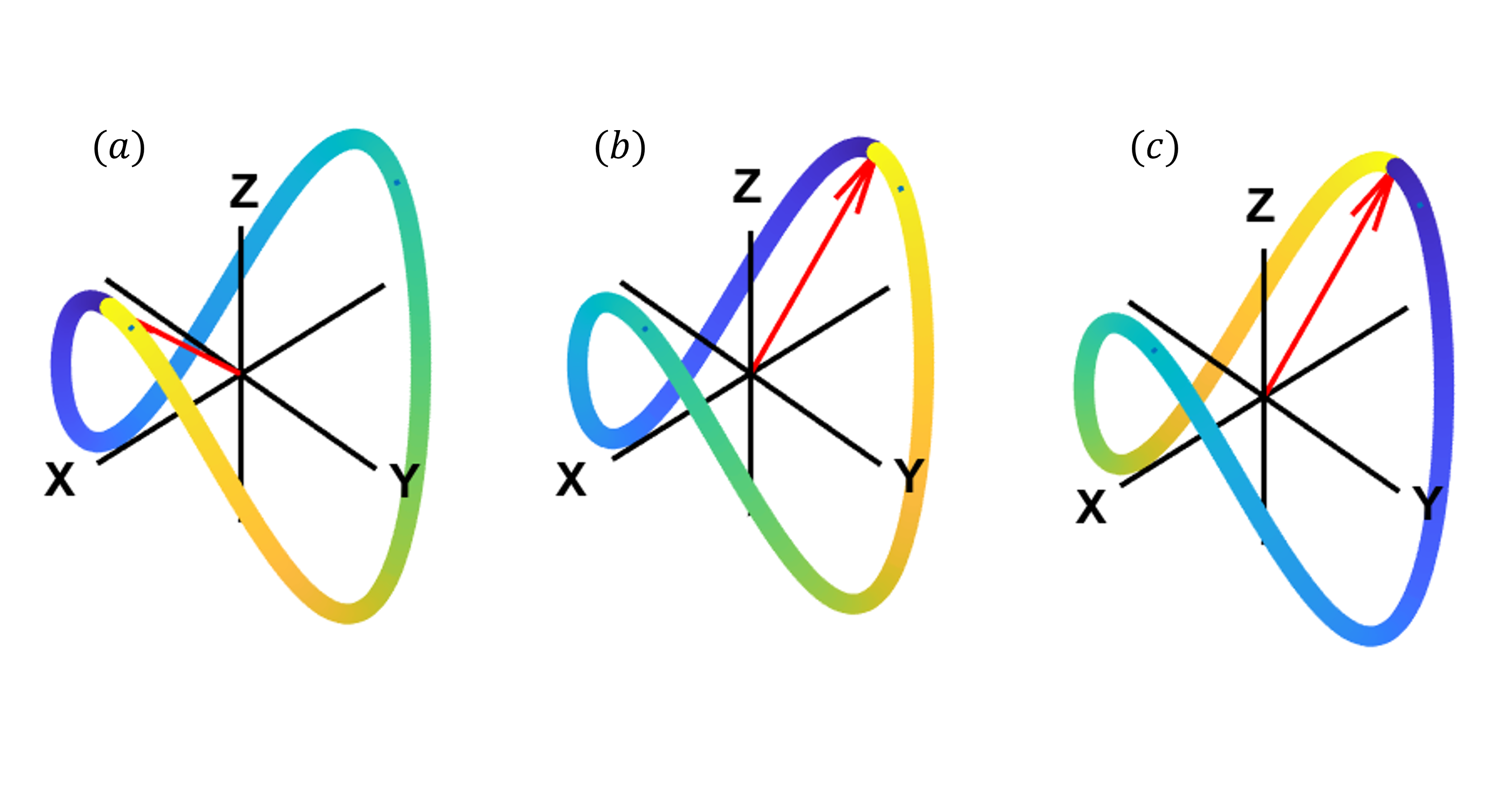}
%% (d)-(f) panels of Bild2.png were referenced in the draft but never discussed;
%% they duplicate Figs. potatoR/InfinityR/InfinityLref/InfinityLT. Re-include if needed:
%\includegraphics[width=0.95\linewidth]{Bild2.png}
\caption{\label{fig:trajs} Orientation trajectories of a rigid symmetric top for the three wavepackets (a) $|\psi\rangle^{L-{\rm Ref}}$, (b) $|\psi\rangle^{R}$, and (c) $|\psi\rangle^{L-\mathcal{T}}$ of \cref{eq:Example1}. Color encodes time as in \cref{fig:potatoR}; the red arrows mark the mean orientation at the same instant $\tau$ in all three panels. Panels (a) and (b) are related by the reflection $\sigma_{YZ}$; panels (b) and (c) by motion reversal; panels (a) and (c) by a rotation by $\pi$ about $Z$.}
\end{figure}
\Cref{fig:trajs}(a--c) shows the orientation trajectories of three simple wavepackets,
 \begin{align}
 |\psi\rangle^{L-{\rm Ref}}(\tau=0) &= |0\,0\,0\rangle + |1\,0\,{-1}\rangle + |2\,0\,{-1}\rangle , \nonumber \\
 |\psi\rangle^{R}(\tau=0) &= |0\,0\,0\rangle + |1\,0\,1\rangle + |2\,0\,1\rangle , \nonumber \\
 |\psi\rangle^{L-\mathcal{T}}(\tau=0) &= |0\,0\,0\rangle - |1\,0\,{-1}\rangle - |2\,0\,{-1}\rangle .
\label{eq:Example1}
\end{align}
The trajectories of $|\psi\rangle^{L-{\rm Ref}}$ and $|\psi\rangle^{R}$ are related by the reflection $\sigma_{YZ}$, those of $|\psi\rangle^{L-\mathcal{T}}$ and $|\psi\rangle^{R}$ by motion reversal (\cref{Tpsi}). The curves are saddle-shaped, bent towards positive $Z$, and as sets of points they are not chiral. They acquire a handedness only when the bend along $Z$ is combined with the sense of rotation in the $XY$ plane, clockwise or counterclockwise. This handedness is local, $h_d(\tau)\neq0$, but both global measures vanish, $\mathcal{C}_g=\mathcal{C}_d=0$, by the symmetry of the curve. Note also that $|\psi\rangle^{L-{\rm Ref}}$ and $|\psi\rangle^{L-\mathcal{T}}$ differ only by the sign of the $M=-1$ components, i.e., by a rotation by $\pi$ about the $Z$ axis: the mirror image and the motion-reversed image of $|\psi\rangle^{R}$ are the same object up to a proper rotation. This is the hallmark of \emph{false} chirality in Barron's sense \cite{Barron1986}. %However it is important to mention that both local chiral measures are non-zero over the whole period and the only reason that the global measures are zero is the symmetry of trajectory. In this case, false chirality in the Barron's sense is not explaining the local dynamical chirality.

On the other hand, this example also illustrates a practical subtlety in comparing enantiomeric wavepackets in a pump--probe experiment. If the rotational wavepacket is created by a short pump pulse at $t=0$ and probed by a short pulse at $t=\tau$, the original wavepacket and its mirror image correspond at the instant $\tau$ to two different molecular orientations (red arrows in \cref{fig:trajs}), and the probe will naturally yield different results for the two, irrespective of chirality. A meaningful enantiosensitive comparison requires either delaying the probe for one of the partners so that both are met at the same orientation, or integrating the probe signal over delays covering the full period $\tau_R$. Likewise, a molecule-fixed transition dipole that is orthogonal to the instantaneous plane of motion spanned by $\mathbf{\Gamma}(\tau)$ and $\dot{\mathbf{\Gamma}}(\tau)$ forms, together with these two vectors, a handed frame; an electronic excitation that is sensitive not only to $\mathbf{\Gamma}(\tau)$ but also to $\dot{\mathbf{\Gamma}}(\tau)$ would then respond with opposite sign to $|\psi\rangle^{R}$ and to $|\psi\rangle^{L-{\rm Ref}}$ or $|\psi\rangle^{L-\mathcal{T}}$.

\subsubsection{A truly chiral example}\label{sec:infinity}
The wavepacket of \cref{fig:InfinityR}, which we take as the right-handed enantiomer $|\psi\rangle^R$, has complex relative phases. Its two left-handed partners, obtained by reflection (case ii) and by motion reversal (case i), are shown in \cref{fig:InfinityLref,fig:InfinityLT}. The reflected partner is the spatial mirror image at every instant, both in the shape of the trajectory and in the direction of motion, while the motion-reversed partner traverses the same curve in the opposite sense. The geometrical and dynamical quantities of the three trajectories are listed in \cref{tab:trajcomp}. As anticipated in \cref{sec:compare}, $\mathcal{C}_g$ changes sign under reflection but not under motion reversal, whereas $\mathcal{C}_d$, $\Omega_s$ and $\Phi_{\rm hol}$ change sign under both. This is obviously a case of true chirality and indeed its relative phases $-e^{2i}$ on $|1\,0\,1\rangle$ and $+1$ on $|1\,0\,{-1}\rangle$ cannot be removed by any rotation about $Z$.

\section{Dependence on the projection $K$}\label{sec:Kdep}
Viewed on the orientation sphere $S^2$, the projected trajectory $\hat{\bm\Gamma}(\tau)$ combines three types of motion, familiar from the classical symmetric top: precession about the conserved angular-momentum direction, along which the trajectory follows a circle of latitude (a great circle for $\theta=\pi/2$); nutation, i.e., a wobbling of the polar angle $\theta$, which turns the latitude circle into a wavy band; and pure precession at fixed $\theta$, for which the trajectory is a small circle.

Since $\langle\boldsymbol\zeta\rangle$ involves only the rank-one components $D^{1*}_{q0}$, the selection rule for the body-fixed projection is $\Delta K=0$. For a symmetric top the $K$ dependence of $E_{JK}$ therefore cancels in all oscillating phases (\cref{B1,B6}): a coherent sum over $K$ does not change the frequencies of the trajectory, but it changes the amplitudes of the different frequency components as well as the time-independent orientation, which enters through the terms proportional to $K$ in \cref{B4,B10}. As a result the trajectory of a wavepacket with $K\neq0$ evolves around a constant orientation displaced from the origin of the space-fixed frame, and with a slightly smaller radius. Classically, increasing $|K|$ narrows the precession cone of the molecular axis about the angular-momentum vector and thus reduces the radius of the trajectory.

The origin of this constant orientation can be understood classically and is derived in \cref{app:permanent}: averaging over the classical precession cancels the transverse component of the molecular axis and leaves $\overline{\boldsymbol{\zeta}}=(J^{cl}_z/(J^{cl})^2)\mathbf{J}^{cl}$, where $\mathbf{J}_{cl}$ is the conserved angular momentum, $J^{cl}=|\mathbf{J}^{cl}|$, and $L_z=\mathbf{J}^{cl}\cdot\boldsymbol{\zeta}$. Its quantum counterpart gives
\begin{align}
    \langle JKM|\zeta_Z|JKM\rangle= \frac{MK}{J(J+1)}
\end{align}
for $J>0$, reproducing the time-independent orientation contributions derived in \cref{B5,B5b,B6} exactly.

\section{Conclusions and prospects}\label{sec:conclusions}
We have introduced chiral rotational wavepackets, coherent superpositions of low-lying rotational states of either chiral or achiral molecule whose mean molecular-axis orientation traces a chiral curve in the laboratory frame, and have characterized their handedness by two derivative-based pseudoscalar measures. The geometrical measure $\mathcal{C}_g$ is an intrinsic, time-even property of the orientation trajectory and provides a sufficient condition for true chirality of the orientation trajectory in Barron's sense. The dynamical measure $\mathcal{C}_d$ is a time-odd, origin-referenced quantity; its local density $h_d$ is proportional to the geodesic curvature of the trajectory projected onto the orientation sphere, which through the Gauss--Bonnet theorem it is related to the solid angle enclosed by the projected trajectory and to the parallel-transport holonomy of tangent vectors parallel transported in the tangent plane along it. Within the single-$K$ setting considered here, wavepackets with real expansion coefficients, such as those generated by a single linearly polarized pump, are only falsely chiral. True chirality requires relative phases between the rotational components that cannot be accounted for by a proper rotation relating the reflected and motion-reversed states. Synthetic chiral light, now available in the optical domain \cite{Greenwood2026,Li2026,Ordonez2026}, imprints exactly such controlled phases between multiphoton pathways and is therefore the natural tool for preparing truly chiral rotational and vibronic wavepackets.

Several extensions are natural. First, the $D^J_{MK}$ functions form a complete basis on the $SO(3)$ manifold and are not restricted to the symmetric top Hamiltonian. Thus the same construction applies to asymmetric tops, to rotors with centrifugal distortion, and to Hamiltonians with additional coupling terms, with the difference that the eigenstates are then linear combinations of $D^{J*}_{MK}$ functions and the orientation trajectory is in general not closed, so that the global measures become long-time averages. Second, the motion-reversed partner of a wavepacket need not be prepared by an actual time reversal: since $\mathcal{T}$ transforms the coefficients according to \cref{Tpsi}, it can be synthesized directly by a shaped pulse, or, for real coefficients, it appears transiently around a rotational revival, where $\Psi(\tau_{\rm rev}-\tau)=\Psi^*(-\tau)$. Third, electronic excitation itself changes the rotational chirality. Excitation by a linearly polarized pulse populates new $K$ components and their coherences, which displaces the orientation trajectory from the origin and breaks the symmetry of the saddle-shaped curves of \cref{fig:trajs}; we expect the resulting Fourier figure to have nonzero $\mathcal{C}_g$, $\mathcal{C}_d$ and holonomy, i.e., photoexcitation can convert a falsely chiral rotational wavepacket into a truly chiral one. This case study will be presented separately. Fourth, and most importantly, the enantiosensitive observables associated with chiral rotational wavepackets remain to be established. A natural setting is the electronic excitation of the rotating molecule by a probe pulse, where a molecule-fixed transition dipole combined with the position and velocity of the mean orientation forms a handed frame; a rotationally invariant pseudoscalar signal, e.g. the difference in the period-integrated absorption of left and right circularly polarized probe light for the two rotational enantiomers, would provide a direct test of the concept, and its connection to photoelectron circular dichroism and related observables is an interesting subject for future work.

\begin{acknowledgments}
Funded by the European Union (ERC, ULISSES, 101054696). Views and opinions expressed are however those of the author(s) only and do not necessarily reflect those of the European Union or the European Research Council. Neither the European Union nor the granting authority can be held responsible for them. We thank Prof.\ M.~Yu.\ Ivanov for reading the manuscript and providing insightful comments and suggestions on its structure and content.
\end{acknowledgments}

\appendix

\section{Matrix representation of the rotation group}\label{app:wignerD}
In the $ZYZ$ convention the rotation matrix $R(\phi,\theta,\chi)$ parametrized by the Euler angles is

\begin{widetext}
\begin{align}\label{eulerMatrix}
&R(\phi,\theta,\chi) =\begin{bmatrix}
\cos\phi\cos\theta\cos\chi - \sin\phi\sin\chi & \sin\phi\cos\theta\cos\chi + \cos\phi\sin\chi  & -\sin\theta\cos\chi \\
-\cos\phi\cos\theta\sin\chi-\sin\phi\cos\chi & -\sin\phi\cos\theta\sin\chi + \cos\phi\cos\chi & \sin\theta\sin\chi \\
\sin\theta\cos\phi & \sin\theta\sin\phi & \cos\theta
\end{bmatrix}
\end{align}
with $0\le\phi<2\pi$, $0\le\theta\leq\pi$, $0\le\chi < 2\pi$. This matrix is also called the direction-cosine matrix. For a vector with Cartesian coordinates $(X,Y,Z)$ in the space-fixed frame and $(x,y,z)$ in the molecular frame,

\end{widetext}
\begin{align}\label{EulerTrans}
    \begin{bmatrix}
x \\ y \\ z
\end{bmatrix} = R(\phi,\theta,\chi)\begin{bmatrix}
X \\ Y \\ Z
\end{bmatrix} .
\end{align}

\subsection{Wigner $D$-matrix}
The Wigner $D$-matrix is an irreducible matrix representation of the rotation group. Its elements (the Wigner $D$-functions) representing the rotation $R(\Omega)$ of $SO(3)$ (integer $J$) are

\begin{align}
    D^{J}_{MK}(\Omega) = \langle J M|R(\Omega)|J K\rangle ,
\end{align}
where $D^{J}_{MK}(\Omega)$ is short for $D^{J}_{MK}(R(\Omega))$. In the Euler-angle parametrization $\Omega = (\phi,\theta,\chi)$,
\begin{align}
    D^{J}_{MK}(\phi,\theta,\chi) = e^{-iM\phi}d^J_{MK}(\theta)e^{-iK\chi},
\end{align}
with $d^J_{MK}(\theta) = \langle JM|e^{-i\theta J_y}|JK\rangle$ the Wigner $d$-functions, i.e., the elements of the reduced rotation matrix \cite{Zare1988,Varshalovich1988}. The relation to the spherical harmonics is
\begin{align}
   & Y_{JM}(\theta,\phi) = \left(\frac{2J+1}{4\pi}\right)^{1/2} D^{(J)*}_{M0}(\phi,\theta,\chi) , \\
   & R(\Omega)Y_{JM}(\theta,\phi) = \sum_{M'}Y_{JM'}(\theta,\phi) D^{(J)}_{M'M}(\Omega) ,
\end{align}
where the first relation is independent of the Euler angle $\chi$. The conjugated elements $D^{J*}_{MK}(\Omega)$ are the elements of the inverse rotation matrix $R^{-1}(\Omega)=R^\dagger(\Omega)$,
\begin{align}\label{Dinver}
D^{J*}_{MK}(\Omega) = [R^{-1}(\Omega)]^{J}_{KM} &= D^{J}_{KM}(-\Omega) \nonumber \\
&= (-1)^{M\,-K} D^{J}_{-M\,-K}(\Omega),
\end{align}
where $\Omega \rightarrow -\Omega$ denotes $(\phi,\theta,\chi) \rightarrow (-\chi,-\theta,-\phi)$. Finally, the integral relation obtained from the Kronecker-product reduction of Wigner $D$-functions, used throughout this work, reads (for integer $J$; see Ref.~\cite{BrownCarrington2003}, Sec.~5.5)
\begin{widetext}
\begin{align}\label{D9}
    \langle J K M | D^{k*}_{pq}(\Omega) | J' K' M'\rangle = (-1)^{M-K} \sqrt{(2J+1)(2J'+1)}
    \begin{pmatrix}
        J & k &J' \\
        -K & q & K'
    \end{pmatrix} \begin{pmatrix}
        J& k & J' \\
        -M & p & M'
    \end{pmatrix}
\end{align}
\end{widetext}
in terms of $3j$-symbols. For $K=K'=0$ it reduces to Gaunt's integral.

\subsection{Irreducible (spherical) tensor representation}
An operator $T^{k} = \{T^{k}_q\}$ with $2k+1$ components ($-k\leq q\leq k$, $k=0, 1 ,2, \dots$) is an irreducible tensor operator of rank $k$ if
\begin{align}
    T'^{k}_q = R\,T^{k}_qR^{-1} = \sum_{p=-k}^kT^{k}_p D^{k}_{pq}(\Omega),
\end{align}
where the left-hand side is the $q$ component of $T^{k}$ in the coordinate system rotated by $R(\Omega)$. Equivalently, using \cref{Dinver},
\begin{align}\label{rotatingtolab}
    T'^{k}_p = R^{-1}T^{k}_pR = \sum_{q=-k}^k D^{k*}_{pq}(\Omega)T^{k}_q .
\end{align}
Under rotations of the coordinate system the spherical tensor components thus transform like the eigenfunctions of the angular-momentum operator, i.e., $T^{k}$ belongs to the irreducible representation $k$ of the rotation group. The (covariant) spherical components of a rank-one tensor (a vector) are related to the Cartesian components by
\begin{align}
    T^{1}_0(\vec{A}) &= A_z , \nonumber \\
    T^{1}_{\pm 1}(\vec{A}) &= \dfrac{\mp 1}{\sqrt{2}} (A_x\pm i A_y),
\end{align}

\section{Expectation values in the rotational wavepacket}\label{app:expect}
To calculate the expectation value of the vector $\boldsymbol{\zeta}$ we need expressions of the form
\begin{widetext}
\begin{align}\label{B1}
    \langle D^{1*}_{10}\rangle  &= \sum_{J'M'K'}\sum_{JMK}b^{J'*}_{M'K'}b^J_{MK} e^{i(E_{J'K'}-E_{JK})\tau}\langle J'M'K'|D^{1*}_{10}|JMK\rangle \nonumber \\
   &= \sum_{J'M'K'}\sum_{JMK}b^{J'*}_{M'K'}b^J_{MK} e^{i(E_{J'K'}-E_{JK})\tau} (-1)^{M'-K'}[(2J+1)(2J'+1)]^{1/2} \nonumber \\
   &  \qquad \times \begin{pmatrix}
        J'& 1 & J \\
        -M' & 1 & M
    \end{pmatrix} \begin{pmatrix}
        J' & 1 &J \\
        -K' & 0 & K
    \end{pmatrix} \left[ \delta_{J',J+1} + \delta_{J',J} +\delta_{J',J-1}\right]\delta_{M',M+1}\delta_{K'K} ,
\end{align}
\end{widetext}
where the Kronecker deltas are redundant and are only written to emphasize which terms survive in the sums. Using the symmetric-top eigenvalues $E_{JK} = J(J+1)+(A/B-1)K^2$, for which the $K$-dependent part cancels in all phases because of $\delta_{K'K}$, we have
\begin{widetext}
\begin{align}
   \langle D^{1*}_{10}\rangle|_{J'=J+1} &= \sum_{JMK}b^{J+1*}_{M+1,K} b^J_{MK} e^{2i(J+1)\tau}  \left[\frac{2(J+M+1)(J+M+2)(J+K+1)(J-K+1)}{(2J+1)(2J+2)^2(2J+3)}\right]^{1/2} , \label{B2} \\
    \langle D^{1*}_{10}\rangle|_{J'=J-1}  &= -\sum_{JMK}b^{J+1}_{M-1,K}b^{J*}_{MK} e^{-2i(J+1)\tau}  \left[\frac{2(J-M+1)(J-M+2)(J+K+1)(J-K+1)}{(2J+1)(2J+2)^2(2J+3)}\right]^{1/2} , \label{B3} \\
   \langle D^{1*}_{10}\rangle|_{J'=J} &= -\sum_{JMK}b^{J}_{MK} b^{J*}_{M+1,K} \frac{K}{J(J+1)} \left[\frac{(J-M)(J+M+1)}{2}\right]^{1/2} . \label{B4}
\end{align}
\end{widetext}
For $\Delta J=-1$ the lower limit of the sum has been set to zero by the relabeling $J-1\rightarrow J$. Using \cref{B2,B3,B4} we obtain
\begin{widetext}
\begin{align}
   \langle  \zeta_X \rangle (\tau) &= \sum_{JMK} \Re\left\{b^{J}_{MK} b^{J*}_{M+1,K}\right\} \frac{K}{J(J+1)}
   \left[(J-M)(J+M+1)\right]^{1/2} \nonumber \\
   & \quad + \Re\left\{- b^{J+1*}_{M+1,K} b^J_{MK} \left[ \frac{(J+M+1)(J+M+2)(J+K+1)(J-K+1)}{(2J+1)(J+1)^2(2J+3)}\right]^{1/2} \right.  \nonumber \\
   &\quad +  \left. b^{J+1}_{M-1,K} b^{J*}_{MK} \left[\frac{(J-M+1)(J-M+2)(J+K+1)(J-K+1)}{(2J+1)(J+1)^2(2J+3)}\right]^{1/2} \right\} \cos(2(J+1)\tau) \nonumber \\
   & \quad + \Im  \left\{ b^{J+1*}_{M+1,K} b^{J}_{MK} \left[\frac{(J+M+1)(J+M+2)(J+K+1)(J-K+1)}{(2J+1)(J+1)^2(2J+3)}\right]^{1/2} \right .\nonumber \\
   & \quad + \left. b^{J+1}_{M-1,K} b^{J*}_{MK} \left[\frac{(J-M+1)(J-M+2)(J+K+1)(J-K+1)}{(2J+1)(J+1)^2(2J+3)}\right]^{1/2}\right\} \sin(2(J+1)\tau) \label{B5}
\end{align}
and
\begin{align}
   \langle \zeta_Y \rangle (\tau) &= \sum_{JMK} \Im\left\{b^{J}_{MK} b^{J*}_{M+1,K}\right\} \frac{K}{J(J+1)}
   \left[(J-M)(J+M+1)\right]^{1/2} \nonumber \\
   & \quad +\Im\left\{- b^{J+1*}_{M+1,K} b^J_{MK} \left[ \frac{(J+M+1)(J+M+2)(J+K+1)(J-K+1)}{(2J+1)(J+1)^2(2J+3)}\right]^{1/2} \right.  \nonumber \\
   & \quad +  \left. b^{J+1}_{M-1,K} b^{J*}_{MK} \left[\frac{(J-M+1)(J-M+2)(J+K+1)(J-K+1)}{(2J+1)(J+1)^2(2J+3)}\right]^{1/2} \right\} \cos(2(J+1)\tau) \nonumber \\
   & \quad - \Re  \left\{ b^{J+1*}_{M+1,K} b^{J}_{MK} \left[\frac{(J+M+1)(J+M+2)(J+K+1)(J-K+1)}{(2J+1)(J+1)^2(2J+3)}\right]^{1/2} \right .\nonumber \\
   & \quad + \left.b^{J+1}_{M-1,K} b^{J*}_{MK} \left[\frac{(J-M+1)(J-M+2)(J+K+1)(J-K+1)}{(2J+1)(J+1)^2(2J+3)}\right]^{1/2}\right\} \sin(2(J+1)\tau) . \label{B5b}
\end{align}
\end{widetext}
For the $\langle \zeta_Z\rangle$ component one needs the expectation value of $D^{1*}_{00}$,
\begin{widetext}
\begin{align}\label{B6}
    \langle D^{1*}_{00}\rangle &= \sum_{J'M'K'}\sum_{JMK}b^{J'*}_{M'K'}b^J_{MK} e^{i(E_{J'K'}-E_{JK})\tau} (-1)^{M'-K'}[(2J+1)(2J'+1)]^{1/2} \nonumber \\
   & \qquad \times \begin{pmatrix}
        J'& 1 & J \\
        -M' & 0 & M
    \end{pmatrix} \begin{pmatrix}
        J' & 1 &J \\
        -K' & 0 & K
    \end{pmatrix} \left[ \delta_{J',J+1} + \delta_{J',J} +\delta_{J',J-1}\right]\delta_{M'M}\delta_{K'K} ,
\end{align}
\end{widetext}
for which
\begin{widetext}
\begin{align}
   \langle D^{1*}_{00}\rangle|_{J'=J+1}   &= \sum_{JMK}b^{J+1*}_{MK} b^J_{MK} e^{2i(J+1)\tau}  \left[\frac{(J+M+1)(J-M+1)(J+K+1)(J-K+1)}{(2J+1)(J+1)^2(2J+3)}\right]^{1/2} , \label{B8}\\
    \langle D^{1*}_{00}\rangle|_{J'=J-1} &= \sum_{JMK}b^{J+1}_{MK}b^{J*}_{MK} e^{-2i(J+1)\tau} \left[\frac{(J+M+1)(J-M+1)(J+K+1)(J-K+1)}{(2J+1)(J+1)^2(2J+3)}\right]^{1/2} , \label{B9}
\end{align}
and, for $J'=J$,
\begin{align}\label{B10}
   \langle D^{1*}_{00}\rangle|_{J'=J} = \sum_{JMK}|b^{J}_{MK}|^2\frac{MK}{J(J+1)} .
\end{align}
The final result is
\begin{align}\label{B11}
   \langle \zeta_Z \rangle (\tau) &= \sum_{JMK} |b^{J}_{MK}|^2 \frac{MK}{J(J+1)} + \left[ \Re\left\{ b^{J+1*}_{MK} b^J_{MK} \right\} \cos(2(J+1)\tau) - \Im\left\{ b^{J+1*}_{MK} b^J_{MK} \right\} \sin(2(J+1)\tau) \right] \nonumber \\
   & \quad \times 2\left[\frac{(J+M+1)(J-M+1)(J+K+1)(J-K+1)}{(2J+1)(J+1)^2(2J+3)}\right]^{1/2} .
\end{align}
\end{widetext}
Note that $K$ enters the time-independent terms of \cref{B5,B5b,B11} linearly, through \cref{B4,B10}, whereas it enters the amplitudes of the oscillating terms only through the even combination $(J+K+1)(J-K+1)$. Changing the sign of $K$ therefore inverts the time-independent orientation offset while leaving the oscillating part of the trajectory unchanged (\cref{tab:effects}).

\section{Fourier figures}\label{app:fourier}
Two-dimensional Lissajous figures are the patterns drawn by two independent perpendicular harmonic oscillators. In analogy with the concept of Fourier knots \cite{Kauffman1998} (self-crossing and open curves are, strictly speaking, not knots), we call a Fourier figure a parametrized curve in $\mathbb{R}^3$ whose three Cartesian coordinate functions are finite Fourier series, i.e., the pattern drawn by three independent oscillators each of which is a linear combination of a finite number of pure frequencies. Three-dimensional Lissajous curves are the subset for which each coordinate function consists of a single harmonic term $A_i\cos(\omega_i \tau+\varphi_i)$. For example, the boundary of a hyperbolic paraboloid (the saddle or "potato chip" of \cref{fig:trajs}) is a three-dimensional Lissajous curve and a trivial knot (unknot), whereas a trefoil is a Fourier figure but not a Lissajous curve.

A Fourier figure is smooth and has finite curvature and torsion. If it is constructed from integer harmonics of a single fundamental frequency it is closed by construction. More generally, Fourier figures with arbitrary frequencies obey the same commensurability rule as Lissajous figures: if all frequencies are rationally related the curve closes; if at least one independent ratio is irrational the curve is quasi-periodic and never closes.

\subsection{Periodicity condition for the orientation trajectory}
The trajectory $\mathbf\Gamma(\tau)$ is closed (periodic) if there exists $\tau_R>0$ such that $\mathbf\Gamma(\tau+\tau_R)=\mathbf\Gamma(\tau)$ for all $\tau$, i.e.,
\begin{align}
\langle \zeta_q\rangle(\tau+\tau_R)=\langle  \zeta_q \rangle(\tau),
\qquad q=0,\pm1 .
\end{align}
For a wavepacket with discrete energies each component is a finite sum of harmonics,
\begin{equation}
\langle \zeta_q\rangle(\tau)=\sum_{\lambda\in\mathcal F_q} A^{(q)}_{\lambda}\,e^{i\omega_{\lambda} \tau},
\end{equation}
where $\mathcal F_q$ is the set of frequencies appearing with nonzero amplitude, determined by the nonzero products of wavepacket coefficients and nonvanishing matrix elements. With $\mathcal F \equiv \mathcal F_{-1}\cup \mathcal F_{0}\cup \mathcal F_{+1}$ the union of all contributing frequencies, the trajectory is closed if and only if there exists $\tau_R>0$ such that
\begin{equation}
e^{i\omega_\lambda \tau_R}=1
\qquad \text{for every } \omega_\lambda\in\mathcal F ,
\end{equation}
i.e., if and only if all contributing frequencies are commensurate, $\omega_\lambda = m_\lambda\omega_0$ with $m_\lambda\in\mathbb Z$ for some $\omega_0>0$. One may then choose $\tau_R=2\pi/\omega_0$, and the minimal period is obtained by taking $\omega_0$ as the greatest common divisor of $\mathcal F$.

For a rigid symmetric top the matrix elements enforce $\Delta K=0$, so only terms with $K'=K$ contribute to $\langle \zeta_q\rangle(\tau)$, and the $K^2$ term of $E_{JK}=J(J+1)+(A/B-1)K^2$ cancels in all phases. All time-dependent contributions therefore oscillate at the frequencies $J'(J'+1)-J(J+1)$, which for the rank-one operator ($\Delta J=\pm1$) are
\begin{equation}
\omega_{J,J+1}=2(J+1), \qquad \omega_{J,J-1}=-2J .
\end{equation}
These are all even integers, so the trajectory is exactly periodic with period $\tau_R=\pi$.

\section{The Frenet frame and the Frenet--Serret formulas}\label{app:TNB}
Let $\boldsymbol{\Gamma}(\tau)$ be a smooth regular curve, $\mathbf{\dot{\Gamma}}:= d\boldsymbol{\Gamma}/d\tau\neq 0$ at every point. Its arc length $s$ is related to the parameter $\tau$ by \cite{Baer2010}
\begin{align}\label{arclength}
\frac{ds}{d\tau} = \Big\|\frac{d\boldsymbol{\Gamma}}{d\tau}\Big\| ,
\end{align}
where $v:=ds/d\tau$ is the speed of the curve, i.e., the magnitude of the velocity of the point $\bm \Gamma$ at time $\tau$. The unit tangent vector at the point $s$ is
\begin{align}\label{Tvec}
\mathbf{T} := \frac{\mathbf{\dot{\Gamma}}}{\|\mathbf{\dot{\Gamma}}\|} = \frac{d\mathbf{\Gamma}}{ds} .
\end{align}
The unit principal normal $\mathbf{N}$ and the curvature $\kappa$ at $s$ are defined by

\begin{align}
&\mathbf{N} :=  \frac{1}{\kappa}\frac{d\mathbf{T}}{ds} , \label{Nvec}\\
& \kappa := \Big\|\frac{d\mathbf{T}}{ds}\Big\| . \label{scurvature}
\end{align}
$\mathbf{N}$ points toward the center of curvature and is perpendicular to $\mathbf{T}$ since $\mathbf{T}\cdot d\mathbf{T}/ds = 0$. The unit binormal

\begin{align}\label{Bvec}
\mathbf{B} := \mathbf{T}\times \mathbf{N}
\end{align}
completes the orthonormal basis. It is perpendicular to the plane spanned by $\mathbf{T}$ and $\mathbf{N}$, the osculating plane, which is the plane that best approximates the curve at $s$. The three vectors form the Frenet (moving) frame, also called the TNB frame \cite{Baer2010,Pressley2010}; the frame twists as it moves along the curve. The local geometry of a curve with a chosen sense of traversal in $\mathbb R^3$ is described by the Frenet--Serret formulas \cite{Baer2010}

\begin{align}
&\frac{d\mathbf{T}}{ds}
= \kappa\mathbf{N} , \label{fs1} \\
&\frac{d\mathbf{N}}{ds}
= -\kappa\mathbf{T} + \alpha\mathbf{B} , \label{fs2} \\
& \frac{d\mathbf{B}}{ds}
= -\alpha\mathbf{N} , \label{fs3}
\end{align}
which follow from \cref{Tvec,Nvec,Bvec} and $\mathbf{N}\cdot d\mathbf{N}/ds = 0$. The parameter $\alpha(s)$ is the torsion of the curve at $s$ and measures how rapidly the osculating plane rotates as one moves along the curve. From \cref{Tvec,fs1,fs2,fs3} the higher derivatives are
\begin{align}\label{highdev}
    \dfrac{d^2\mathbf{\Gamma}}{ds^2} &= \kappa \mathbf{N} , \\
    \dfrac{d^3\mathbf{\Gamma}}{ds^3} &= \dfrac{d\kappa}{ds} \mathbf{N} -\kappa^2 \mathbf{T} +\kappa \alpha \mathbf{B} .
\end{align}

\section{Relation between the $H^{(3)}$ correlation function and the dynamical measure}\label{app:H3}
The chiral correlation function $H^{(3)}$ of Refs.~\cite{Ayuso2019,Ayuso2022} is the triple product of the vector at three consecutive times $\tau_0$, $\tau_0+\Delta\tau $ and $ \tau_0 + 2\Delta\tau$. Let $\bm \Gamma_0:=\bm \Gamma(\tau_0)$ and
\begin{widetext}
\begin{align}\label{eq:expansion}
    \bm \Gamma_1 &= \bm \Gamma(\tau_0+\Delta\tau )
    = \bm \Gamma_0 +\Delta\tau \, \mathbf{\dot{\Gamma}}_0 + \frac{\Delta\tau^2}{2}\mathbf{\ddot{\Gamma}}_0 +  \frac{\Delta\tau^3}{6}\mathbf{\dddot{\Gamma}}_0 +\mathcal{O}(\Delta\tau^4) \equiv \bm \Gamma_0 + \Delta \bm \Gamma_1 , \nonumber \\
     \bm \Gamma_2 &= \bm \Gamma(\tau_0+2\Delta\tau )
     = \bm \Gamma_0 +2\Delta\tau \, \mathbf{\dot{\Gamma}}_0 + 2\Delta\tau^2\mathbf{\ddot{\Gamma}}_0 +  \frac{4\Delta\tau^3}{3}\mathbf{\dddot{\Gamma}}_0 +\mathcal{O}(\Delta\tau^4) \equiv \bm \Gamma_0 + \Delta \bm \Gamma_2 ,
\end{align}
where the subscript $0$ denotes evaluation at $\tau_0$. Then
\begin{align}
    H^{(3)}_{\rm loc}(\tau_0;\Delta\tau) & = \bm \Gamma_0\cdot [(\bm \Gamma_0 + \Delta \bm \Gamma_1)\times (\bm \Gamma_0+\Delta \bm \Gamma_2)] \nonumber \\
    &= \bm \Gamma_0\cdot (\bm \Gamma_0\times \Delta \bm \Gamma_2) + \bm \Gamma_0 \cdot (\Delta \bm \Gamma_1 \times \bm \Gamma_0) + \bm \Gamma_0 \cdot (\Delta \bm \Gamma_1\times \Delta \bm \Gamma_2) \nonumber \\
    &=\bm \Gamma_0 \cdot [\Delta\tau^3(\mathbf{\dot{\Gamma}}_0 \times \mathbf{\ddot{\Gamma}}_0 )+ \mathcal{O}(\Delta\tau^4)] ,
\end{align}
\end{widetext}
where the first two terms vanish identically and, in the last one, the contributions $2\Delta\tau^3\,\mathbf{\dot{\Gamma}}_0\times\mathbf{\ddot{\Gamma}}_0$ and $-\Delta\tau^3\,\mathbf{\dot{\Gamma}}_0\times\mathbf{\ddot{\Gamma}}_0$ combine. Hence
\begin{align}\label{eq:H3loc}
    \lim_{\Delta\tau\rightarrow 0} \frac{H^{(3)}_{\rm loc}(\tau_0;\Delta\tau)}{\Delta\tau^3} &= \bm \Gamma(\tau_0) \cdot (\mathbf{\dot{\Gamma}}(\tau_0) \times \mathbf{\ddot{\Gamma}}(\tau_0)) \nonumber\\ &= \|\mathbf{\dot\Gamma}(\tau_0)\|^3\,h_d(\tau_0) ,
\end{align}
which is \cref{eq:H3limit}. For a unit-speed parametrization, $\|\mathbf{\dot\Gamma}\|=1$, the average of $H^{(3)}_{\rm loc}/\Delta\tau^3$ over the trajectory therefore equals $\mathcal{C}_d$, \cref{eq:Cd}; for a general parametrization the two differ by the weight $\|\mathbf{\dot\Gamma}\|^{2}$ inside the integral.

\section{The dynamical measure as a geodesic curvature}\label{app:geodesic}
Let $S$ be an oriented surface with outward unit normal $\mathbf{n}$, and $\mathbf{\Gamma}$ a unit-speed curve ($\|\mathbf{\dot\Gamma}\|=1$) on $S$. The vectors $\mathbf{\dot\Gamma}$, $\mathbf{n}$ and $\mathbf{n} \times \mathbf{\dot\Gamma}$ form an orthonormal basis, the Darboux frame. Since $\mathbf{\ddot\Gamma}\perp \mathbf{\dot{\Gamma}}$,
\begin{align}\label{F1}
    \mathbf{\ddot\Gamma} = \kappa_n \mathbf{n} + \kappa_g (\mathbf{n} \times \mathbf{\dot{\Gamma}}) ,
\end{align}
where $\kappa_n$ and $\kappa_g$ are the normal and geodesic curvatures \cite{Pressley2010}, related to the curvature $\kappa$ of the curve by $ \kappa^2 = \kappa_n^2 + \kappa_g^2$. From \cref{F1},
\begin{align}\label{F2}
    \kappa_g = \mathbf{\ddot\Gamma} \cdot (\mathbf{n}\times\mathbf{\dot\Gamma}) = \mathbf{n}\cdot (\mathbf{\dot\Gamma}\times \mathbf{\ddot\Gamma}) .
\end{align}
On the unit sphere $S^2$ the normal at a point is the position vector itself, hence
\begin{align}\label{F3}
     \kappa_g = \mathbf{\Gamma}(\tau)\cdot (\mathbf{\dot\Gamma}(\tau) \times \mathbf{\ddot\Gamma}(\tau)) ,
\end{align}
valid for unit-speed curves on the unit sphere, $\| \mathbf{\Gamma}(\tau)\| = 1$.

The geodesic curvature measures how much a curve bends within the surface, away from a geodesic; it is a geometric property of the curve and does not depend on how fast a point travels along it. From \cref{F2}, $\kappa_g$ changes sign when the orientation of either the curve ($\dot{\bm\Gamma}$) or the surface ($\mathbf{n}$) is reversed \cite{Pressley2010}; in particular, reversing the sense of traversal for a fixed orientation of the surface changes the sign of $\kappa_g$.

We now consider the parallel transport of a vector along the curve. The covariant derivative of a vector $\mathbf{X}$ tangent to $S^2$ is \cite{Baer2010}
\begin{align}
    \dfrac{\nabla \mathbf{X}}{ds} := \dfrac{d\mathbf{X}}{ds}-\Big(\dfrac{d\mathbf{X}}{ds}\cdot \mathbf{n}\Big)\mathbf{n} .
\end{align}
Since $\mathbf{X}$ is tangent to the sphere it can be written in the Darboux basis as $\mathbf{X} = a_X \,\mathbf{\dot{\Gamma}} + b_X \, (\mathbf{n}\times\mathbf{\dot{\Gamma}}) $. Using \cref{F1} and the relation
\begin{align}
    \frac{d[\mathbf{n}\times\mathbf{\dot{\Gamma}}] }{ds} = -\kappa_g \mathbf{\dot{\Gamma}} ,
\end{align}
valid for curves on the unit sphere (where $d\mathbf{n}/ds=\mathbf{\dot\Gamma}$ and $\kappa_n = -1$ for the outward normal), one finds
\begin{align}
     \dfrac{\nabla \mathbf{X}}{ds}  = \Big(\dfrac{da_X}{ds} - b_X\,\kappa_g\Big) \mathbf{\dot{\Gamma}} + \Big(a_X\,\kappa_g + \dfrac{db_X}{ds} \Big) (\mathbf{n}\times\mathbf{\dot{\Gamma}}) .
\end{align}
Parallel transport, $ \nabla \mathbf{X}/ds =0$, then gives for the angle $\varphi:= \tan^{-1}(a_X/b_X)$ between $\mathbf{X}$ and the tangent vector
\begin{align}
     \dfrac{d\varphi}{ds}  = -\kappa_g ,
\end{align}
so that after one traversal of the closed curve the parallel-transport holonomy is
\begin{align}\label{F8}
     \Phi_{\rm hol} = \Delta\varphi  = -\oint_\Gamma \kappa_g ds \pmod{2\pi} .
\end{align}
If one is interested only in the evolution of the molecular-axis direction, independently of the degree of orientation, the projected trajectory on the unit sphere and its geodesic curvature thus provide the appropriate description; the degree of orientation does not contribute to the holonomy of \cref{F8}, which is caused solely by the rotation of the tangent plane along the curve.

For the general orientation trajectory of \cref{curve}, which is neither normalized nor of unit speed, \cref{F3} must be applied to the projected curve $\hat{\bm\Gamma}=\bm\Gamma/\|\bm\Gamma\|$ reparametrized by its own arc length, $ds=\|d\hat{\bm\Gamma}/d\tau\|\,d\tau$. Writing $\hat{\bm\Gamma}=\bm\Gamma/r$ with $r=\|\bm\Gamma\|$, the derivatives $\dot{\hat{\bm\Gamma}}$ and $\ddot{\hat{\bm\Gamma}}$ differ from $\dot{\bm\Gamma}/r$ and $\ddot{\bm\Gamma}/r$ only by terms proportional to $\bm\Gamma$ and $\dot{\bm\Gamma}$, which drop out of the triple product with $\hat{\bm\Gamma}$, so that $\hat{\bm\Gamma}\cdot(\dot{\hat{\bm\Gamma}}\times\ddot{\hat{\bm\Gamma}})=\bm\Gamma\cdot(\dot{\bm\Gamma}\times\ddot{\bm\Gamma})/r^3$. Dividing by $\|\dot{\hat{\bm\Gamma}}\|^3$ to account for the arc-length parametrization gives
\begin{align}\label{F9}
     \kappa_g = \frac{\bm\Gamma\cdot(\dot{\bm\Gamma}\times\ddot{\bm\Gamma})}{\|\bm\Gamma\|^3\,\|\dot{\hat{\bm\Gamma}}\|^3}
     = \left(\dfrac{\|\mathbf{\dot\Gamma}\|}{\|\mathbf{\Gamma}\| \, \| \dot{\hat{\bm\Gamma}}\|} \right)^3 h_d ,
\end{align}
which relates the geodesic curvature of the projected trajectory to the local dynamical measure of \cref{dmeasure}.

\section{Classical origin of permanent molecular orientation}
\label{app:permanent}

Consider a torque-free rigid symmetric top with principal moments
$I_x=I_y=I_\perp$ and $I_z$ about its molecular symmetry axis
$\boldsymbol{\zeta}$. The angular momentum $\mathbf{J}^{cl}$ and its
body-axis projection $J^{cl}_z=\mathbf{J}^{cl}\cdot\boldsymbol{\zeta}$ are
conserved. For $J^{cl}=|\mathbf{J}^{cl}|>0$, the axis therefore precesses
uniformly about $\mathbf{J}^{cl}$ at a fixed angle $\theta_0$, with
$\cos\theta_0=J^{cl}_z/J^{cl}$. Separating its parallel and perpendicular
components gives
\begin{equation}
\boldsymbol{\zeta}(t)
=\cos\theta_0\,\frac{\mathbf{J}^{cl}}{J^{cl}}
+\sin\theta_0\,\mathbf{e}_\perp(t),
\end{equation}
where $\mathbf{e}_\perp(t)$ is a unit vector rotating in the plane
perpendicular to $\mathbf{J}^{cl}$. Averaging over one precession cycle
eliminates this transverse component, leaving
\begin{equation}\label{eq:classical-permanent-orientation}
\overline{\boldsymbol{\zeta}}
=\frac{J^{cl}_z}{(J^{cl})^2}\mathbf{J}^{cl},
\qquad
\overline{\zeta_Z}=\frac{J^{cl}_zJ^{cl}_Z}{(J^{cl})^2},
\end{equation}
where $J^{cl}_Z$ is the laboratory-$Z$ projection.
This surviving component is the classical origin of persistent
symmetric-top orientation~\cite{Xu2023}. A linear rotor has $J^{cl}_z=0$,
so its cycle-averaged orientation vanishes.

The corresponding quantum substitutions $(J^{cl})^2\rightarrow\hbar^2J(J+1)$,
$J^{cl}_z\rightarrow\hbar K$, and $J^{cl}_Z\rightarrow\hbar M$ yield
$\overline{\langle\boldsymbol{\zeta}_Z\rangle} =  MK/[J(J+1)]$, which is also the exact expectation value of
$\zeta_Z$ of a pure $|JKM\rangle$ state  in Eq.(\ref{B6}).
More generally, within a fixed $(J,K)$ manifold the orientation
operator is proportional to the angular-momentum operator, with
factor $K/[\hbar J(J+1)]$. Defining the wavefunction component corresponding to a given J and K,
$|\Psi_{JK}\rangle=\sum_M b^J_{MK}|JKM\rangle$, we therefore obtain the time-independent orientation of quantum mechanical symmetric top
\begin{equation}\label{eq:wavepacket-permanent-orientation}
\overline{\langle\boldsymbol{\zeta}\rangle}
=\sum_{J=1}^{\infty}\sum_{K=-J}^{J}
\frac{K}{\hbar J(J+1)}
\langle\Psi_{JK}|\widehat{\mathbf{J}}|\Psi_{JK}\rangle .
\end{equation}
Coherences between degenerate $M$ levels within each $(J,K)$
manifold remain stationary and can support transverse orientation. The $J=0$ state contributes zero. Permanent orientation thus reflects the locking of the precession-averaged molecular axis to the conserved angular-momentum direction.

\bibliographystyle{apsrev4-2}
\bibliography{refs}
\end{document}